\documentclass[fleqn,usenatbib]{mnras}

\usepackage{newtxtext,newtxmath}
\usepackage[T1]{fontenc}

\DeclareRobustCommand{\VAN}[3]{#2}
\let\VANthebibliography\thebibliography
\def\thebibliography{\DeclareRobustCommand{\VAN}[3]{##3}\VANthebibliography}

\usepackage{graphicx}	% Including figure files
\usepackage{amsmath}	% Advanced maths commands
\usepackage{xcolor}     % Edit text colour
\usepackage{adjustbox}  % Adjust table size
\usepackage{stfloats}
\usepackage{makecell}
\usepackage{jabbrv}
\DefineJournalAbbreviation{Monthly Notices of the Royal Astronomical Society}{MNRAS}

\title[FLARES XVII: Galaxy-halo connection]{First Light and Reionisation Epoch Simulations (FLARES) XVII: Learning the galaxy-halo connection at high redshifts}

\author[M. G. A. Maltz et al.]
{Maxwell G. A. Maltz,$^{1}$\thanks{E-mail: m.maltz@sussex.ac.uk}
Peter A. Thomas,$^{1}$
Christoper C. Lovell,$^{2}$
William J. Roper,$^{1}$\newauthor
Aswin P. Vijayan,$^{1}$
Dimitrios Irodotou,$^{3}$
Shihong Liao,$^{4}$
Louise T. C. Seeyave,$^{1}$
and Stephen M. Wilkins$^{1}$
\\
$^{1}$Astronomy Centre, University of Sussex, Falmer, Brighton BN1 9QH, UK\\
$^{2}$Institute of Cosmology and Gravitation, University of Portsmouth, Burnaby Road, Portsmouth PO1 3FX, UK\\
$^{3}$The Institute of Cancer Research, 123 Old Brompton Road, London SW7 3RP, UK\\
$^{4}$Key Laboratory for Computational Astrophysics, National Astronomical Observatories, Chinese Academy of Sciences, Beijing 100101,
China}

\date{Accepted XXX. Received YYY; in original form ZZZ}

\pubyear{2024}

\begin{document}
\label{firstpage}
\pagerange{\pageref{firstpage}--\pageref{lastpage}}
\maketitle

\begin{abstract}
Understanding the galaxy-halo relationship is not only key for elucidating the interplay between baryonic and dark matter, it is essential for creating large mock galaxy catalogues from $N$-body simulations. High-resolution hydrodynamical simulations are limited to small volumes by their large computational demands, hindering their use for comparisons with wide-field observational surveys.
We overcome this limitation by using the First Light and Reionisation Epoch Simulations (\textsc{Flares}), a suite of high-resolution ($M_{\rm{gas}} = 1.8\ \times\ 10^6\ \rm{M}_\odot$) zoom simulations drawn from a large, $(3.2\ \rm{cGpc})^3$ box.
We use an extremely randomised trees machine learning approach to model the relationship between galaxies and their subhaloes in a wide range of environments.
This allows us to build mock catalogues with dynamic ranges that surpass those obtainable through periodic simulations.
The low cost of the zoom simulations facilitates multiple runs of the same regions, differing only in the random number seed of the subgrid models; changing this seed introduces a butterfly effect, leading to random differences in the properties of matching galaxies. This randomness cannot be learnt by a deterministic machine learning model, but by sampling the noise and adding it post-facto to our predictions, we are able to recover the distributions of the galaxy properties we predict (stellar mass, star formation rate, metallicity, and size) remarkably well.
We also explore the resolution-dependence of our models’ performances and find minimal depreciation down to particle resolutions of order $M_{\rm{DM}} \sim 10^8\ \rm{M}_\odot$, enabling the future application of our models to large dark matter-only boxes.
\end{abstract}

\begin{keywords}
galaxies: haloes -- galaxies: high-redshift -- methods: data analysis -- methods: numerical -- cosmology: large-scale structure of Universe -- software: machine learning
\end{keywords}

%%%%%%%%%%%%%%%%%%%%%%%%%%%%%%%%%%%%%%%%%%%%%%%%%%

%%%%%%%%%%%%%%%%% BODY OF PAPER %%%%%%%%%%%%%%%%%%

\section{Introduction}
In the standard $\Lambda$CDM model, the gravitational collapse of pressureless dark matter formed the first bound structures: \textit{haloes}. These haloes, some of which contain virialised substructures, termed \textit{subhaloes}, provided the gravitational wells in which galaxies would later form (\citealt{white_1978}; \citealt{blumenthal_1984}; \citealt{davis_1985}), seeding the origins of a close relationship between the properties of galaxies and their host haloes, known as the \textit{galaxy-halo connection}; see \citealt{wechsler_galaxy-halo_connection} for a complete review.

Simulations that probe the large-scale structure (LSS) of the Universe need to be sufficiently large -- of order $\sim 1\ \rm{Gpc}$ in length -- to sample the rare density fluctuations that seeded the growth of the first massive galaxies (\citealt{bagla_2006}; \citealt{cui_2018}; \citealt{lovell_first_2020}). Hydrodynamical simulations are the most self-consistent tool for testing our theoretical understanding of the galaxy-halo connection, but the complex physics involved, which requires significant computational cost, means they are typically limited to box sizes of length $\sim 100\ \rm{Mpc}$ (e.g. \citealt{schaye_eagle_2015}; \citealt{springel_2018}; \citealt{dave_2019}). They are prohibitively expensive to run on Gpc-scales without restrictive compromises being made in resolution or the redshift at which they are stopped. For example, the recent FLAMINGO simulations \citep{flamingo_schaye_2023} are the largest hydrodynamical simulations -- ranging from $1$ to $2.8\ \rm{cGpc}$ in box length -- ever run to $z=0$, but the low resolutions (gas particle masses of order $10^8$ to $10^9\ \rm{M}_\odot$) place a prohibitive lower limit on the (sub)structures that can be resolved. Consequently, the number and quality of the galaxies available for analysis is hindered. Additionally, there are far fewer satellite galaxies resolved, which affects clustering statistics on scales $\leq 1\ \rm{Mpc}$ \citep{wechsler_galaxy-halo_connection}. The BlueTides simulation \citep{bluetides}, meanwhile, has length $\sim 740\ \rm{cMpc}$, and a similar resolution to EAGLE \cite[gas particle mass of $2.4 \times 10^6\ \rm{M}_\odot$]{schaye_eagle_2015}, but was only run to $z=7$, prohibiting comparisons with local observations.

Typically, the LSS is studied using $N$-body simulations, comprising collisionless dark matter only (e.g. \citealt{springel_2005}). The removal of baryonic physics makes dark matter-only (DMO) simulations much cheaper and quicker to run, but comparisons with observations necessarily rely on an existing understanding of the galaxy-halo connection. Such an understanding allows $N$-body simulations to act as skeletons onto which galaxies can be painted. Several methods for doing this exist, ranging from the more empirical (e.g. halo occupation distributions) to the more physical (e.g. semi-analytic models). We briefly review some of these methods here.

Halo occupation distributions (HODs) empirically calculate the probability distribution for the number of galaxies satisfying some criteria, such as minimum stellar mass, for a halo of given mass – or as a function of multiple halo properties (\citealt{Zheng_2005}; \citealt{hearin_2013}). Mock galaxy catalogues can then be created using Monte Carlo methods.

Subhalo abundance matching (SHAM) models assume a monotonic relationship between galaxy and subhalo masses, typically associating the most massive galaxies with the most massive subhaloes \citep{vale_linking_2004}. Other properties, such as galaxy luminosity \citep{simha_testing_2012} and the subhaloes' maximum rotational speed \citep{chaves-montero_subhalo_2016}, may be used as substitutes for mass, but all SHAM approaches rely on a simple rank ordering.

The drawback of both HODs and SHAMs is their dependence on just one or a limited number of subhalo properties. Despite the strong correlation between galaxy and subhalo masses, there is a significant amount of scatter in the properties of galaxies at any fixed subhalo mass. This scatter arises from \textit{galaxy assembly bias} -- the dependence of galaxy properties on subhalo properties other than mass \citep{zentner_2014} -- as well as mergers, environmental effects, and feedback processes, all of which are neglected by traditional HOD and SHAM approaches (\citealt{Lehmann_2017}; \citealt{Hadzhiyska_2020}).

Semi-analytic models (SAMs) apply approximate, analytical relationships between
galaxies and their host haloes in order to map galaxies onto \textit{N}-body simulations. SAMs incorporate a wide range of physical processes such as gas cooling, star formation, supernova feedback, and AGN feedback (\citealt{somerville_1999}; \citealt{Croton_2006}; \citealt{Guo_2011}; \citealt{henriques_2015}). They are, therefore, much more physically motivated than the HOD and SHAM approaches, but they require the exploration of a vast parameter space, involving extensive calibration against observations\footnote{A problem also suffered by hydrodynamical simulations when calibrating subgrid models.}.

Recently, machine learning (ML) algorithms, including tree-based algorithms (e.g. \citealt{kamdar_machine_2016_hydro}; \citealt{jo_machine-assisted_2019}), generalised additive models (e.g. \citealt{Hausen_2023}), and neural networks (e.g. \citealt{moster_2021}; \citealt{jespersen_textttmangrove_2022}; \citealt{chittenden_2023}), have been applied to the learning of the galaxy-halo connection through the combined use of hydrodynamical and DMO simulations, allowing the properties of any galaxy to be predicted given a set of inputs characterising its halo or subhalo. The ML models are trained to reproduce the outputs of existing hydrodynamical simulations, which are more complex and physical than SAMs, but once trained, the application of ML models to DMO simulations demands less computational power than SAMs and considerably less than the running of hydrodynamical simulations of equivalent volume \citep{jespersen_textttmangrove_2022}. Furthermore, ML models are able to extract complex, non-linear relationships from the data themselves, dispensing the need for pre-defined functional forms and other simplifying assumptions.

One particularly useful class of simulations in the application of ML to the galaxy-halo connection are zoom simulations (\citealt{lovell_machine_2021}; \citealt{deAndres_2023}). Zoom simulations take small regions from a large $N$-body simulation and resimulate them at high resolution with full hydrodynamics. By choosing which regions to resimulate, a much larger range of environments can be studied than would occur within a periodic hydrodynamical simulation of comparable volume, thus allowing for the simulation of many more massive galaxy clusters and protoclusters (\citealt{barnes_cluster-eagle_2017}; \citealt{lovell_first_2020}). Due to their small volumes, zoom simulations cannot themselves be used to study the LSS. However, by learning the relationships between galaxies in the high-resolution resimulations and their haloes in corresponding DMO simulations, the distribution and properties of galaxies across large $N$-body boxes can be predicted and used to generate mock catalogues, at a fraction of the computational cost of running hydrodynamical simulations of comparable size. This opens the pathway to the use of high-resolution hydrodynamical simulations in the study of large-scale structure. Furthermore, once a machine learning model has been trained, the predictive power of the input features can be examined (\citealt{moster_2021}; \citealt{desanti_mimicking_2022}; \citealt{Mcgibbon_2023}). This insight into which halo features are most useful in predicting galaxy properties can be used to inform semi-analytic models, as well as HODs and SHAMs that rely on halo features beyond mass. This is one example of the complementarity that exists between different approaches to generating mock catalogues.

ML methods are not without their drawbacks, however. The predictions from ML models are often biased towards the peak of the true distributions, leading to an underprediction of the scatter in galaxy-halo relationships such as the stellar-halo mass relation (SHMR). Attempts to overcome this issue, such as oversampling rarer galaxies \citep{desanti_mimicking_2022}, and using predicted galaxy properties as inputs in a second model \citep{hernandez_not_2023}, have enjoyed moderate success at mimicking the true distributions. However, deterministic ML models trained only on DM features have been consistently unable to recover the full extent of the scatter in galaxy-halo relationships \citep{stiskalek_scatter_2022}.

In this study, we train an extremely randomised trees (ERT) model to predict four key galaxy properties\footnote{In keeping with common machine learning parlance, we will refer to the data we aim to emulate (the
galaxy properties) as the \textit{target variables}, and the input data on
which the machine is trained (the dark matter halo properties) as the
\textit{features}.} -- stellar mass, time-averaged star formation rate (SFR)\footnote{Time-averaged over $100\ \rm{Myr}$.}, metallicity, and stellar half-mass radius -- in the First Light and Reionisation Epoch Simulations (\textsc{Flares}) (\citealt{lovell_first_2020}; \citealt{vijayan_first_2020}) suite of 40 zoom simulations at two redshifts: $z=5$ and $z=10$. We exploit the cheap nature of zoom simulations by running additional versions of two \textsc{Flares} regions with different random number seeds in the subgrid models. Changing the seed creates a butterfly effect, resulting in significant differences between like-for-like galaxies (\citealt{keller_chaos_2019}; \citealt{borrow_2023}). This places a fundamental limit on how well a deterministic ML model trained on DM features can perform. We sample scatter from the distribution of differences between like-for-like galaxies and add it to our ML predictions post-facto. This enables us to better replicate the true galaxy distributions.

This paper is structured as follows. Section \ref{sec:simulations} introduces the simulations used and the method for matching subhaloes between simulations; we also use this section to assess the scatter in galaxy properties in different runs of the same simulations. In Section \ref{sec:ml}, we introduce the machine learning and data-processing methods employed, as well as our chosen features and target variables. We then present the results of our models tested on the fiducial-resolution DMO simulations in Section \ref{sec:results_eagle-res}. Finally, we apply our models to several lower-resolution DMO simulations in Section \ref{sec:results_low-res}. This provides insight into the accuracies obtainable when applying our models to different $N$-body boxes. Concluding remarks are presented in Section \ref{sec:conclusions}.
\label{sec:intro}

\section{Simulations}
\label{sec:simulations}

In this section, we introduce the \textsc{Flares} suite of hydrodynamical zoom simulations, and the DMO simulations that we use for our model training. We then explore how much of the variance in galaxy properties at fixed subhalo mass is numerical.

\subsection{First Light and Reionisation Epoch Simulations}
\label{sec:flares}

The First Light and Reionisation Epoch Simulations (\textsc{Flares}) (\citealt{lovell_first_2020}; \citealt{vijayan_first_2020}) are a suite of 40 zoom simulations comprising spherical regions of radius $14\ h^{-1}\ \rm{cMpc}$, drawn from the same $(3.2\ \rm{cGpc})^3$ parent simulation as the C-EAGLE simulations \citep{barnes_cluster-eagle_2017}. The simulations were run down to $z=4.7$ with snapshots saved at integer redshifts $z \in [5,15]$, thus covering the entire epoch of reionisation (EoR) -- the period in which the first stars and galaxies were born, returning the gas content of the Universe to the ionised state in which it formed \citep{BARKANA2001125}. The full range of environments present in the parent simulation were sampled, including the 16 most overdense and two most underdense regions (see Table A1 from \citealt{lovell_first_2020} for a list of the overdensities of each region). The focus on overdense environments means that, at $z=5$, \textsc{Flares} samples $\sim 20$ times more $10^{10}\ \rm{M}_\odot$ galaxies than the EAGLE simulations, and extends the maximum galaxy mass from $\sim 10^{10.5}\ \rm{M}_\odot$ to $\sim 10^{11.3}\ \rm{M}_\odot$, despite the total resimulated volume being less than $50$ per cent larger than the EAGLE box. A weighting scheme, whereby each region is weighted according to the frequency
with which its environment appears in the parent simulation, allows statistical
distribution functions to be produced for the whole parent volume. These distributions have been shown to match a number of observationally-derived distribution functions remarkably well (\citealt{lovell_first_2020}; \citealt{vijayan_first_2020}; \citealt{roper_first_2022}; \citealt{Donnan2024}).

\textsc{Flares} uses the AGNdT9 parameter configuration of the EAGLE model \citep{schaye_eagle_2015}, the same resolution as EAGLE (gas particle mass $M_{\rm{gas}} = 1.8 \times 10^6\ \rm{M}_\odot$ and dark matter particle mass $M_{\rm{DM}} = 9.7 \times 10^6\ \rm{M}_\odot$), a softening length of $2.66\ \rm{ckpc}$, and a  Planck year 1 cosmology ($\Omega_0 = 0.307$; $\Omega_\Lambda = 0.693$; $h=0.6777$) \citep{planck_collaboration_planck_2014}.

\subsection{Dark matter-only simulations}
\label{sec:dmo}

In order to apply our models to $N$-body boxes, the models must be trained on features that have been extracted from DMO simulations. Baryonic effects in hydrodynamical simulations alter the properties of their dark matter haloes (\citealt{duffy_impact_2010}, \citealt{castro_impact_2020}), meaning that models trained on haloes from hydrodynamical simulations cannot be applied to DMO simulations without introducing biases. Instead, we must match galaxies from the hydrodynamical simulations to their haloes from corresponding DMO simulations of the same regions.

The \textsc{Flares} DMO parent simulation was chosen for its large volume, which contains a distribution of overdensities representative of the whole Universe. However, its resolution ($M_{\rm{DM}} = 8.01\ \times\ 10^{10}\ \rm{M}_{\odot}$) is too low for halo-finding algorithms to identify all but the most massive structures. It is, therefore, ill-suited for learning the galaxy-halo connection. We opt to resimulate all 40 \textsc{Flares} regions with dark matter only\footnote{It should be noted that DMO simulations contain the Universal mass content of baryons, but treat all particles as collisionless.} at the fiducial EAGLE resolution ($M_{\rm{DM}} = 1.15 \times 10^7\ \rm{M}_\odot$)\footnote{The corresponding hydrodynamical simulations split this mass into equal numbers of gas and dark matter particles with mass resolutions corresponding to their relative abundances: $16$ per cent and $84$ per cent of $M_{\rm{DM}} = 1.15 \times 10^7\ \rm{M}_\odot$, respectively.}. The machine learning (ML) analysis conducted on these fiducial-resolution regions is discussed in Section \ref{sec:results_eagle-res}. We then apply our ML models trained at the fiducial resolution to three lower-resolution DMO simulations, whose resolutions are obtained by dividing the total number of particles, $N$, that would exist in the parent volume (if run at fiducial resolution) by factors of $8$, $64$, and $512$, corresponding to particle masses of $M_{\rm{DM}} = 8.92 \times 10^7\ \rm{M}_{\odot}$, $M_{\rm{DM}} = 7.14 \times 10^8\ \rm{M}_{\odot}$, and $M_{\rm{DM}} = 6.52 \times 10^9\ \rm{M}_{\odot}$, respectively\footnote{The softening length in the lower-resolution simulations scales by $\left(\frac{M_{\rm{DM,low-res}}}{M_{\rm{DM},fiducial\;res}}\right)^{1/3}$.}. We refer to these resolutions as '$N/8$', '$N/64$', and '$N/512$.' For the purposes of this investigation, we select a single region to re-run at lower resolutions. We choose the most overdense region (Region 00), which has a central overdensity\footnote{$\delta \equiv (\rho/\overline{\rho}) - 1$, where $\rho$ is the density of the region, smoothed over a spherical volume of radius $14\ h^{-1}\ \rm{cMpc}$, and $\overline{\rho}$ is the universal mean density.} of $\delta = 0.970$ at $z=4.67$ \citep{lovell_first_2020}, thus guaranteeing that the models are exposed to a large range of subhalo masses, including those that form the most massive protoclusters. We examine how well our models generalise to these lower resolutions in Section \ref{sec:results_low-res}. This provides crucial insight into how well our models could be expected to perform when applied to different $N$-body simulations.

\subsection{Structure finding}

Haloes are found using a friends-of-friends (FoF) algorithm. Substructure within the haloes is identified using \textsc{Subfind} \citep{springel_populating_2001}. Subhaloes are catalogued only if they contain 20 bound particles or more. However, subhaloes towards this particle limit may be spurious detections, and a higher threshold of $\sim 100$ particles is typically needed to obtain reliable measurements of subhalo properties \citep{onions_subhaloes_2012}. To ensure robustness, we therefore apply a mass cut of $M_{\rm{subhalo}} \geq 10^{10.5}\ \rm{M}_{\odot}$ to the DMO-simulation subhaloes included in our model training and analysis. This means that each fiducial-resolution DMO subhalo contains at least 2,750 particles. A $10^{10.5}\ \rm{M}_{\odot}$ subhalo contains 355 and 44 particles at $N/8$ and $N/64$ resolution, respectively, while the minimum subhalo mass possible at $N/512$ resolution is $10^{11.1}\ \rm{M}_{\odot}$. We note that the properties of some subhaloes at the lowest resolutions may not be well constrained. However, it is our objective to examine the extent to which this decrease in the resolution of subhaloes affects the performance of our machine learning models.

In most instances, the FoF haloes are not resolved into smaller substructures by \textsc{Subfind}. However, all model training is performed on \textsc{Subfind} `subgroups', thus making use of substructure where it exists. In the discussion that follows, we use the word \textit{subhalo} to refer to such structures, irrespective of whether or not they are resolved elements of FoF haloes.

\subsection{Subhalo matching}
\label{section:halo_matching}

We reiterate that, even when run at identical resolutions on the same initial conditions, the properties of subhaloes in corresponding hydrodynamical and DMO simulations differ, partly due to numerical effects, but mostly due to the effects of baryonic physics on the dark matter distribution. This necessitates the formulation of a method by which subhaloes in the different simulations can be matched.

\begin{figure}
    \includegraphics[width=0.5\textwidth]{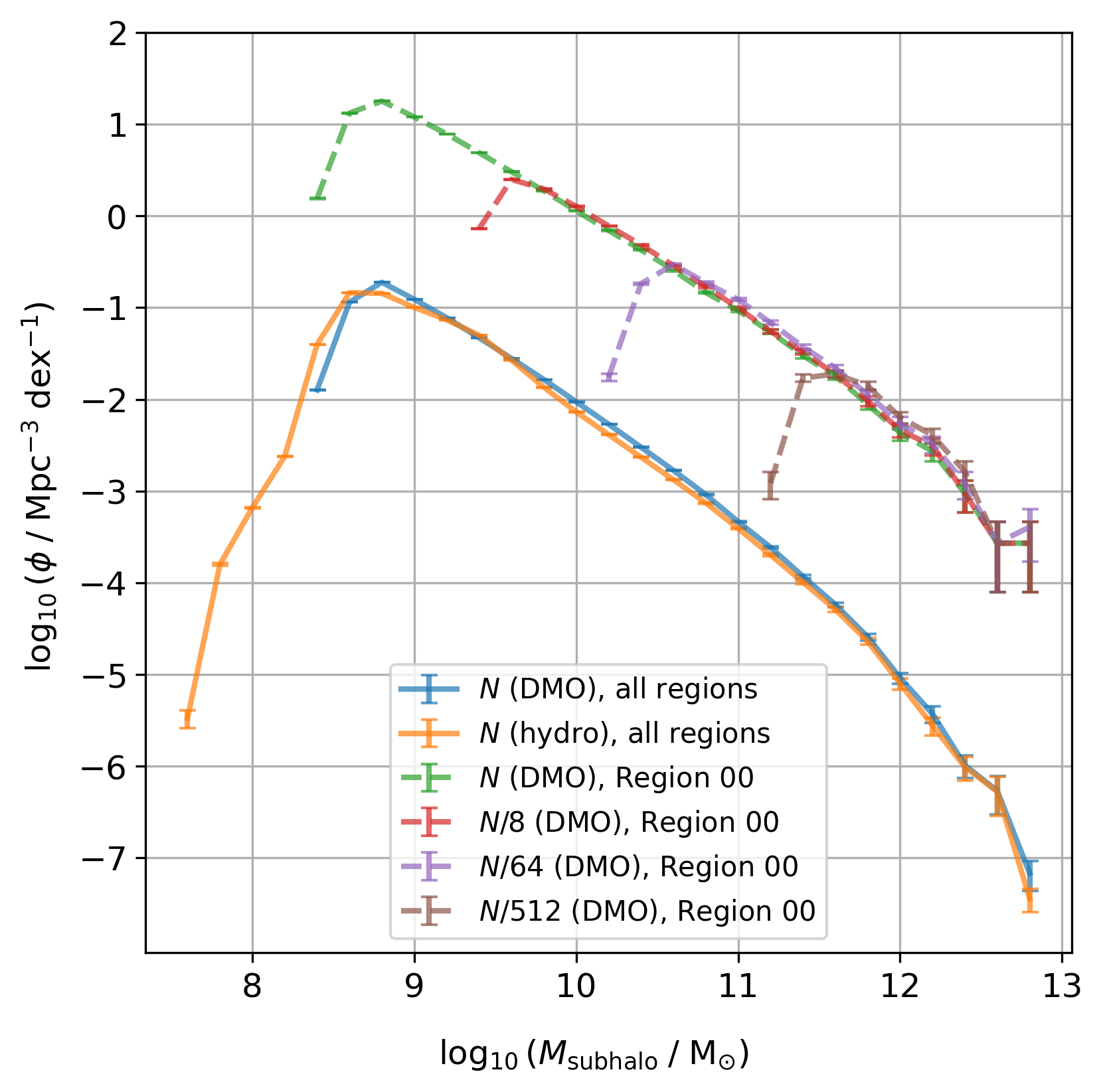}
	\caption{Subhalo mass functions (SHMFs) at $z=5$, shown for the 40 resimulations comprising the \textsc{Flares} suite (solid lines; weighting scheme applied), run with dark matter-only (blue) and full hydrodynamics (orange), and for the most overdense region, Region 00 (dashed lines; weighting scheme not applied), run with dark matter-only at four different resolutions. These resolutions are denoted $N/k$, where $k$ is the factor by which the number of particles, $N$, that would exist in a fiducial-resolution run of the parent simulation, is reduced. Error bars show the $1\sigma$ Poisson uncertainties in the number counts.}
	\label{fig:subhalo_number_densities}
\end{figure}

In Figure \ref{fig:subhalo_number_densities}, we show the subhalo mass function (SHMF) at $z=5$ for each of the simulations -- or suite of simulations -- that we discuss. The SHMF, $\phi(M_{\rm{subhalo}})$, traces the number of subhaloes per unit volume, $n$, per mass interval, as a function of subhalo mass, $M_{\rm{subhalo}}$:
\begin{equation}
    \label{eqn:mass_function}
    \phi(M_{\rm{subhalo}}) = n(M_{\rm{subhalo}})\;/\;\rm{Mpc}^{-3}\ \rm{dex}^{-1}
\end{equation}

The \textsc{Flares} suite is biased towards extreme regions, so we apply weights to each region to get the Universal SHMF. The scheme used is outlined in Section 2.4 of \citet{lovell_first_2020}. SHMFs at $z=5$ are shown with this weighting scheme applied in Figure \ref{fig:subhalo_number_densities}. This figure also shows SHMFs at the same redshift for the simulations of Region 00 run at lower resolutions. For these SHMFs, the weighting scheme is turned off to illustrate how many more subhaloes this region contains compared with the Universal mean.

Since \textsc{Subfind}'s resolution limit is defined by particle count, the inclusion of lower-mass baryonic particles in the hydrodynamical simulations results in a lower mass limit for resolved subhaloes than in the DMO simulations of equivalent resolution; this explains why the orange tail extends to lower subhalo masses than the blue tail in Figure \ref{fig:subhalo_number_densities}.

Naturally, the amount of substructure identified increases with resolution. It is also the case that massive subhaloes in a DMO simulation may be split into multiple smaller subhaloes in a hydrodynamical simulation of equivalent resolution. This is evident in Figure \ref{fig:subhalo_number_densities}, which shows a very slightly lower normalisation for the hydrodynamical SHMF than that of its DMO counterpart. This is a problem when it comes to matching subhaloes by the bijective commonality of their most-bound particles \citet{lovell_machine_2021}. We overcome this problem by matching each hydrodynamical subhalo with the highest-mass DMO subhalo that is in its vicinity. We argue that the former is likely to be the 'same' structure as the latter \textit{or} a resolved element of it, particularly in the high-mass regime we are interested in.

We build a K-dimensional (KD) tree of the centre of potentials (COPs) of subhaloes in each DMO simulation. KD trees are data structures used to the find points in different datasets that have the closest matching values \citep{maneewongvatana_analysis_2002}. In this case, $\rm{K}=3$, corresponding to three spatial dimensions.

We link the $i^{\rm{th}}$ subhalo from the DMO simulation to the $j^{\rm{th}}$ subhalo from the hydrodynamical simulation if 
\begin{equation}
    \Delta R_{ij} \le k \times (R_{1/2,\rm{hydro},\mathit{i}}\ +\ R_{1/2,\rm{DMO},\mathit{j}}),
\end{equation}
where $R_{1/2,\rm{hydro},\mathit{i}}$ and $R_{1/2,\rm{DMO},\mathit{j}}$ are the subhaloes' half-mass radii, and $k$ is a coefficient whose optimal value we discuss below. If a hydrodynamical-simulation subhalo is linked with more than one DMO-simulation subhalo, we match it with the one of greatest mass.

Because there may be multiple hydrodynamical subhaloes matched to any given DMO subhalo, the mass of each hydrodynamical subhalo matched to the $i^{\rm{th}}$ DMO subhalo is summed. We repeat this process on a sample of regions for different half-integer values of $k$. For each $k$, we calculate a weighted root mean squared deviation (RMSD) between the logged masses of the DMO subhaloes and the summed masses of their hydrodynamical matches, defined as
\begin{equation}
    \resizebox{\linewidth}{!}{$\mathrm{Weighted}\;\mathrm{RMSD} =  \sqrt{\frac{\sum_{\mathit{i}=1}^{\mathit{N}}w_i\left(\log_{10}\left(M_{\rm{subhalo}\;\rm{DMO},\mathit{i}}\;/M_{\rm{hydro}\;\rm{matches},\mathit{i}}\right)\right)^2}{\mathit{N}}}$},
\end{equation}
where $N$ is the number of matches and the $w_i$ are weights. To calculate the weights, we split the range of DMO subhalo masses into 10 equal-width bins in logspace. The weight given to a subhalo in any given bin is then the inverse of the number count of subhaloes in that bin. For a given resolution and redshift, we choose the value of $k$ that produces the lowest weighted RMSD among subhaloes in the DMO simulation with masses $M_{\rm{subhalo}} \geq 10^{10.5}\ \rm{M}_{\odot}$. Thus, we prioritise the accurate matching of rare, high-mass subhaloes. The optimal value of $k$ is found to be $1.0$ at fiducial resolution, $1.5$ at $N/8$ and $N/64$ resolution, and $2.0$ at $N/512$ resolution. This increase in the optimal $k$ with decreasing resolution arises from our reduced ability to accurately resolve the COPs of the DMO subhaloes at reduced resolution.

\begin{figure}
    \includegraphics[width=0.5\textwidth]
    {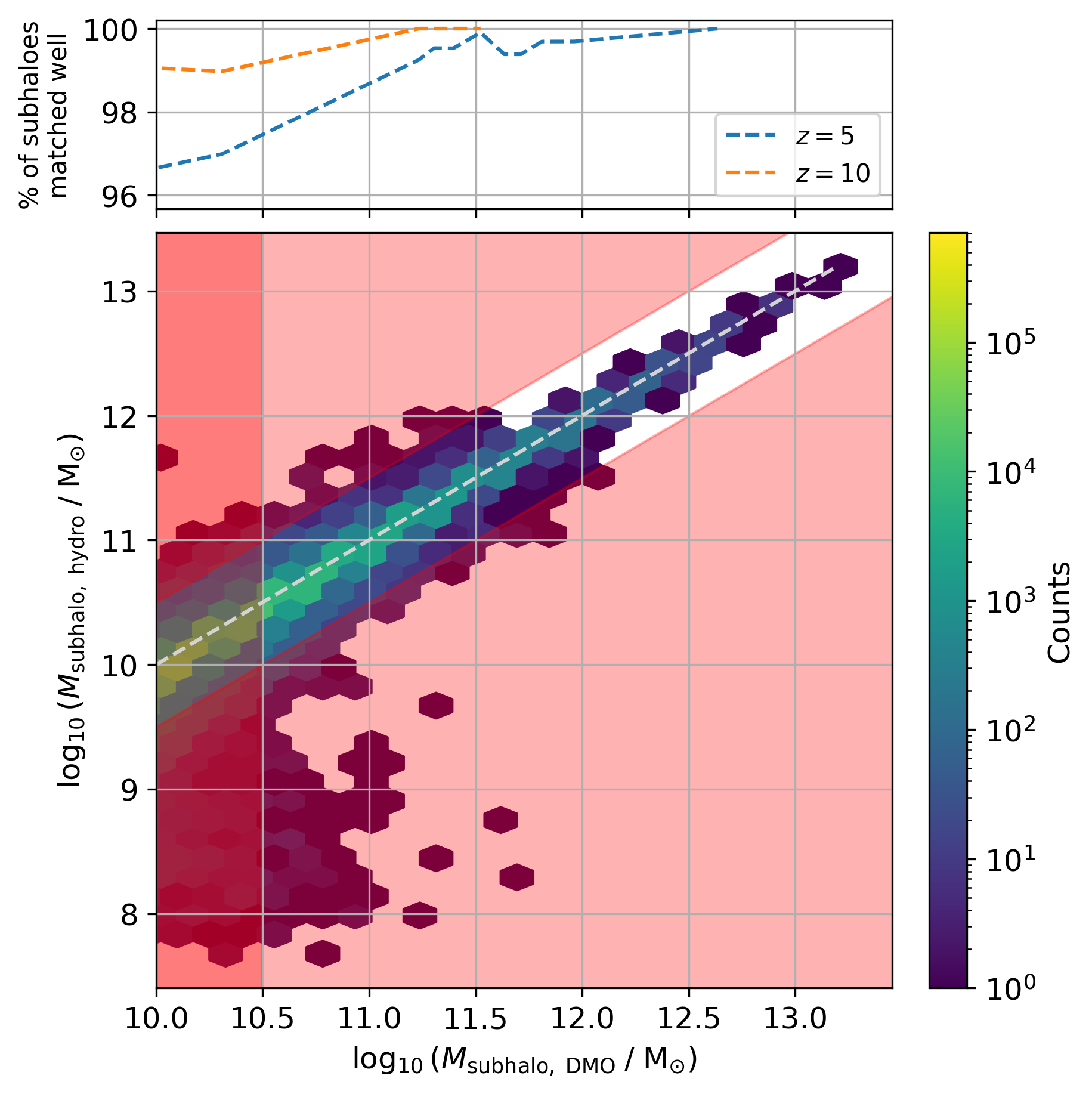}
    \caption{Subhalo matching for all 40 \textsc{Flares} regions run at fiducial resolution.
    Top panel: the percentage of subhaloes in the DMO simulations that are 'matched well' to subhaloes from the hydrodynamical simulations, where subhaloes that are 'matched well' have masses that differ by $\leq 0.5$ dex. Where more than one hydrodynamical subhalo has been matched to a DMO subhalo, their masses have been summed.
    Bottom panel: mass-mass plot of matching hydrodynamical and DMO subhaloes, combining data from $z=5$ and $z=10$.
    DMO subhaloes with $M<10^{10.5}\ \rm{M}_{\odot}$, and whose masses differ by more than 0.5 dex from those of their hydrodynamical matches, are discarded from the datasets used for machine learning. The regions of the plot that contain discarded data are shaded red.}
    \label{fig:masses_halo_matches}
\end{figure}

The bottom panel of Figure \ref{fig:masses_halo_matches} shows a mass-mass plot, in hexbins, of subhalo matches in the fiducial-resolution simulations for $z=5$ and $z=10$ combined. The $y$-axis shows the sum of the masses of the hydrodynamical subhaloes that match to each DMO subhalo. The majority of matched subhaloes have masses that straddle the one-to-one line, particularly at the highest masses, indicating that they are well-matched.

The top panel of Figure \ref{fig:masses_halo_matches} shows the percentage of DMO subhaloes matched to hydrodynamical subhaloes whose combined masses are within $\pm 0.5$ dex of their own. We discard subhaloes whose matches lie outside of this range. This is done to minimise the impact of 'bad' matches -- typically, cases where some components of the subhalo are missing in the hydrodynamical simulation, likely due to disruption from baryonic feedback. In doing this, we discard no more than 0.8 per cent of subhaloes from our fiducial-resolution catalogue, so this pruning is unlikely to introduce any significant biases. The fraction of 'good' matches is $\gtrsim 97.5$ per cent for all $M_{\rm{subhalo}} \geq 10^{10.5}\ \rm{M}_\odot$, and increases with increasing mass. It's also larger at $z=10$ than $z=5$, most probably due to the time-dependent growth of the impact of baryonic processes on the subhalo dynamics. Figure \ref{fig:mass_ratio_matches} in Appendix \ref{sec:appendix_mass_ratios} shows how the mass of the most massive hydrodynamical subhalo matched to each DMO subhalo compares with the masses of any other matches. We find that, apart from at the lowest resolution ($N/512$), the mass of the most massive match usually dominates the combination of all other matches.

Once subhaloes have been matched, galactic properties -- by their nature only present in hydrodynamical simulations -- can be assigned to subhaloes in the DMO simulations. We do this as follows: for stellar mass and SFR, we take the sum of the quantities in the matching hydrodynamical subhaloes; for metallicity, we take the average value in the matching subhaloes, weighted by their mass; for the stellar half-mass radius (HMR), we take the stellar HMR of the most massive matching subhalo. Since the most massive match typically dwarfs the masses of the others, this is a reasonable approximation to make.

Table \ref{table:fid-res_halos} details the number of DMO subhaloes above our mass cut of $10^{10.5}\ \rm{M}_{\odot}$ at $z=5$ and $z=10$, combined across all 40 regions of the fiducial-resolution simulations, and the fraction of those that we include in our datasets for machine learning. The same statistics for the lower-resolution simulations (Region 00 only) are presented in Table \ref{table:low-res_halos}.

\begin{table}
\begin{tabular}{lllll}
                                                     & $z=5$           & $z=10$          &  &  \\ \cline{1-3}
DMO subhaloes with $M \geq 10^{10.5}\ \rm{M}_{\odot}$ & 43,304          & 958           &  &  \\
... Matched to hydro subhaloes...                        & 43,167 (99.7\%) & 957 (99.9\%)   &  &  \\
... Masses of matches within 0.5 dex...                 & 42,950 (99.2\%) & 956 (99.8\%)  &  &  \\
... Matches contain galaxies...                         & 42,950 (99.2\%) & 956 (99.8\%)  &  &  \\
... Matches contain star formation ...                  & 42,249 (97.6\%) & 956 (99.8\%) &  & 
\end{tabular}
\caption{The total number of DMO subhaloes across all 40 regions with masses $M \geq 10^{10.5}\ \rm{M}_{\odot}$ (first row); the number of these DMO subhaloes that match to at least one hydrodynamical subhalo (second row); the number of these DMO subhaloes whose hydrodynamical matches have a combined mass that lies within a range of 0.5 dex either side of their mass (third row); the number of these DMO subhaloes whose matching hydrodynamical subhaloes contain at least one star particle (fourth row); the number of these DMO subhaloes whose matching hydrodynamical subhaloes contain a non-zero star formation rate (fifth row). In each cell, the percentage in brackets denotes the percentage of DMO subhaloes with $M_{\rm{subhalo}} \geq 10^{10.5}\ \rm{M}_{\odot}$ that fulfill the criteria up to and including that row. It can be seen that we discard less than 1 per cent of all subhaloes in this mass regime.}
\label{table:fid-res_halos}
\end{table}

\subsection{Intrinsic stochasticity}
\label{sec:intrinsic scatter}

We now explore the stochasticity inherent to the hydrodynamical simulations.
Several previous works have shown that cosmological simulations display chaotic behaviour, whereby small perturbations in the initial conditions grow into large macroscopic differences at later times (\citealt{thiebaut_onset_2008}; \citealt{genel_quantification_2019}; \citealt{keller_chaos_2019}; \citealt{borrow_2023}). For example, \citet{genel_quantification_2019} found that perturbing the initial positions of particles in two otherwise identical hydrodynamical simulations\footnote{Using the IllustrisTNG model.} by order $\sim 10^{-7}\ \rm{pc}$ led to differences in the stellar masses of matched galaxies of $\sim12$ per cent by $z=0$. These differences were found to be larger in simulations that incorporated feedback processes than those that did not.
This is because feedback is itself a source of chaotic behaviour. Cosmological simulations are, typically, only resolved down to kpc scales. The effect on the larger environment of processes that occur at sub-kpc scales -- such as star formation and black hole accretion -- must be approximated using subgrid models. These subgrid models often use pseudo-random numbers to determine the occurrence -- or lack thereof -- of events such as star formation and supernovae at a given time step. Changing the random number seed creates a ripple effect that, over time, results in macroscopic differences between two pairs of otherwise identical simulations (\citealt{keller_chaos_2019}; \citealt{borrow_2023}).
This inherent randomness places a limit on how well deterministic machine learning models can perform. After all, they can only learn from the dark matter features they are provided.

To remedy this limitation, we learn to mimic the stochasticity by modelling its dependence on subhalo features. To do so, we run additional hydrodynamical simulations of two \textsc{Flares} regions\footnote{Regions 00 and 24, which have central overdensities of $\delta = 0.970$ and $-0.007$, respectively (c.f. Table A1, \citealt{lovell_first_2020}). We choose to investigate two regions with very different overdensities to check that the statistics of the differences in galaxy properties are independent of the large-scale environment.} with different random number seeds to those in the original runs. Upon matching galaxies from these simulations to subhaloes in the corresponding DMO simulations, we obtain two realisations of galaxy properties for each DMO subhalo. By modelling the distribution of differences in the properties of like-for-like galaxies, we can randomly sample appropriate amounts of scatter to add to our model predictions, enabling us to better recover the overall distributions of galaxy properties. We demonstrate this in Section \ref{sec:adding_scatter}.

\begin{figure*}
    \centering
    \includegraphics[width=0.9\linewidth]{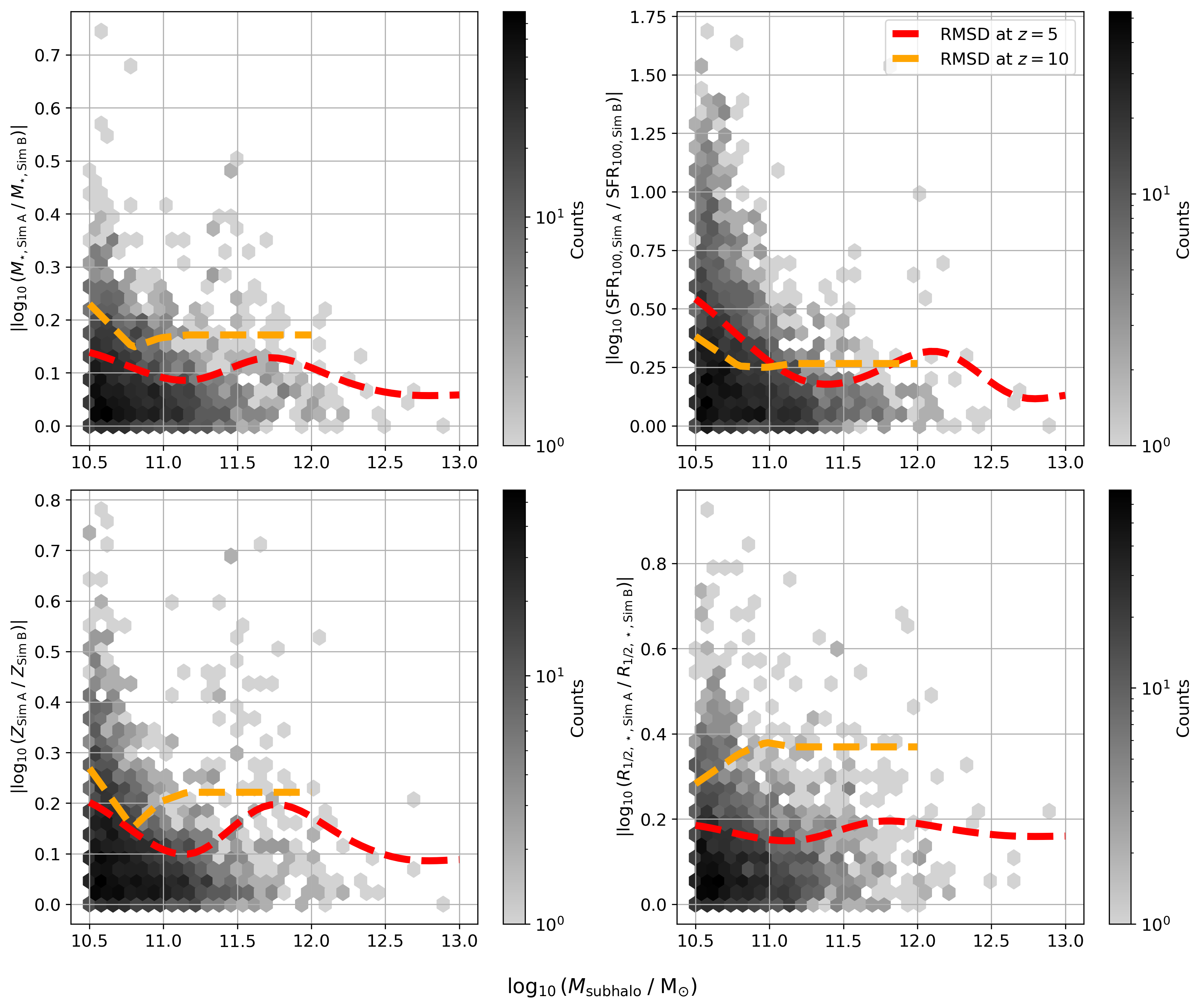}
    \caption{Differences between the Sim A and Sim B realisations of stellar mass, time-averaged SFR, metallicity, and stellar half-mass radius, shown as a function of subhalo mass. The hexbins show the absolute logspace differences for the $z=5$ and $z=10$ snapshots combined. The red and orange dashed lines show spline fits to the RMSD in bins of subhalo mass for $z=5$ and $z=10$, respectively.}
    \label{fig:rmsd_rand_seed}
\end{figure*}

For each of the alternative random number seed runs, we repeat the subhalo matching process outlined in Section \ref{section:halo_matching}, and assign galaxy properties to the DMO subhaloes in the same way. We denote the two sets of galaxy properties assigned to the DMO subhaloes as coming from 'Sim A' and 'Sim B.' The absolute differences between the Sim A and Sim B values are shown in hexbins as a function of subhalo mass in Figure \ref{fig:rmsd_rand_seed}.
We also show the root mean squared deviation (RMSD) between the two realisations, separately for $z=5$ and $z=10$:
\begin{equation}
    \rm{RMSD} =    \sqrt{\frac{\sum_{\mathit{i}=1}^{\mathit{N}}\left(\rm{Sim\;A}_{\mathit{i}} - \rm{Sim\;B}_{\mathit{i}}\right)^2}{\mathit{N}}}.
\end{equation}    
Here, $\rm{Sim\;A}_{\mathit{i}}$ and $\rm{Sim\;B}_{\mathit{i}}$ are the $i^{\rm{th}}$ instance of the logged value in Sim A and Sim B, respectively. For each galaxy property, we calculate the RMSD associated with a subhalo by considering the 10 nearest subhaloes in log mass. To smooth out the noise, we then compute the mean RMSD in five equal-width bins in logspace and fit spline curves to these values. For $z=5$, we fit cubic splines; for $z=10$, where there are far fewer data, we fit linear splines to avoid overfitting. We only include DMO subhaloes whose masses are within $\pm 0.5\ \rm{dex}$ of their matches in Sim A \textit{and} Sim B.

Figure \ref{fig:rmsd_rand_seed} shows that stellar mass is the property least sensitive to changes in the random number seed. The spread of differences in SFR\footnote{We calculate the SFR time-averaged over $100\ \rm{Myr}$. We do this by finding the age and masses of all star particles bound to the galaxy. We then sum the masses of those with ages $\leq\ 100\ \rm{Myr}$, and normalise the total by $100\ \rm{Myr}$.} at fixed subhalo mass is greatest, reaching over 1 dex in several instances. This can be attributed to the bursty and stochastic nature of star formation at high redshifts, which is especially prominent in lower-mass galaxies \citep{sun_seen_2023}. The redshift-dependence of feedback intensity and variability is also reflected by typically larger RMSDs at $z=10$ than $z=5$. This suggests that, although the only difference between the two simulation set-ups is a numerical change in the subgrid models, the differences at resolved scales are manifested -- at least partly -- through physical processes. The role played by numerics on resolved scales could be quantified by re-running the hydrodynamical simulations at lower resolutions and measuring the increase in differences between like-for-like galaxies.

At $z=5$, prominent bumps in the RMSD of stellar mass and metallicity are observed at $M_{\rm{subhalo}} \sim 10^{11.7}\ \rm{M}_\odot$, and in the RMSD of SFR at $M_{\rm{subhalo}} \sim 10^{12.1}\ \rm{M}_\odot$\footnote{The existence of this bump is independent of our binning method.}. We posit that these bumps arise from AGN feedback. Black holes residing in the centres of subhaloes with $10^{11.5} \leq M_{\rm{subhalo}} \leq 10^{12.5}\ \rm{M}_\odot$ undergo rapid growth \citep{mcalpine_link_2017}, the feedback from which induces stochasticity in the properties of the host galaxy. \citet{piana_2022} found that the efficiency of AGN feedback peaks at $M_{\rm{subhalo}} \sim 10^{11.75}\ \rm{M}_\odot$ at $z=4$, remarkably close to the subhalo masses at which the RMSDs peak at $z=5$ in Figure \ref{fig:rmsd_rand_seed}. At lower subhalo masses, the central black holes are too small to undergo significant accretion, whereas the stronger gravitational potential at higher subhalo masses restricts the heating and expelling of gas.
At $z=10$, there are too few subhaloes with $M_{\rm{subhalo}} \geq 10^{11.5}\ \rm{M}_\odot$ to determine the presence -- or lack thereof -- of similar bumps. However, given a large enough sample size, we would expect any such bumps to be less prominent due to the reduced time available for black hole growth.

Finally, we emphasise that, although the properties of individual galaxies matched to the same subhaloes differ in simulations that use different random number seeds, the physics used in each simulation is exactly the same, and so the global summary statistics of the two galaxy populations are consistent.

\section{Machine Learning}
\label{sec:ml}

We now introduce our machine learning pipeline that will be used to generate our initial predictions, which we later augment by adding intrinsic scatter.

\subsection{Extremely Randomised Trees}
\label{sec:ert}

We opt to use the Extremely Randomised Trees (ERT) \citep{geurts_extremely_2006} regressor, which has been used extensively throughout astronomy literature and has been shown to recover simulated galaxy properties to high precision (\citealt{kamdar_machine_2016_hydro}; \citealt{jo_machine-assisted_2019}; \citealt{lovell_machine_2021}). We implement our ERT models using the \textsc{Scikit-learn} Python package \citep{scikit-learn}.
Like the also-popular random forest (RF) algorithm, ERT uses an ensemble of decision trees. However, there are some differences between the two methods. Both incorporate safeguards against overfitting -- the common issue in ML practices in which an algorithm can learn to predict the data it is trained on \textit{too well}, to the extent that it is poor at generalising to unseen data -- but they do so in different ways. Decision trees in RFs are built using random sub-samples of the training dataset. At each node of a decision tree, the feature and threshold that produce the most optimal split among the bootstrapped sample are chosen (for example, $M_{\rm{subhalo}} > 10^{11}\ \rm{M}_\odot$). The downside of using optimal splits at the decision nodes is that dominant features and patterns in the dataset can lead to the forest consisting of many similar-looking trees. An ensemble of similar-looking trees would be sensitive to the exact values of the training data, and may therefore be poor at generalising. RFs reduce the occurrence of similar-looking trees by bootstrapping the training data. Trees in ERT models, on the other hand, are built on the entire dataset, but use random thresholds at each node. The introduction of this randomisation means no two trees are the same, making ERTs slightly less prone to variance than RFs. The predicted output of the ERT model is an average of the predictions made across each decision tree in the ensemble.

\subsection{Features and target variables}
\label{sec:features}

We simultaneously train on data from two redshifts, $z=5$ and $z=10$, allowing the models to learn the redshift-dependence of the galaxy-halo connection\footnote{Other snapshots can easily be incorporated, which is something we plan to do in future work as a necessary preliminary to generating mock lightcones.}.

We aim to predict four key galaxy properties: \textbf{stellar mass}, \textbf{star-formation rate (SFR)} (averaged over 100 Myr\footnote{We predict a time-averaged SFR as this smoothes out any short-term fluctuations associated with the instantaneous SFR, while making the data more comparable with observational results \citep{donnari_2019}.}), \textbf{metallicity} (defined as the mass fraction of the galaxy's star particles that comprise elements heavier than helium), and \textbf{size} (the half-mass radius (HMR) of the galaxy's stellar component). We choose these properties due to their importance in creating mock galaxy observables (\citealt{guidi_2016}; \citealt{tang_2021}). These galaxy properties will be referred to as our target variables.
For stellar mass, metallicity and stellar HMR, we use the values returned by \textsc{Subfind}. The time-averaged SFR is calculated by finding the age and masses of all bound star particles.

To predict the target variables, we train the machine to learn their dependence on the properties of their dark matter subhalo and surrounding environment. We refer to the input data we use as the features. Rather than throwing all possible features at the machine, we train it to learn the relationships we expect to be physically meaningful. Arbitrarily increasing the number of features can lead to the \textit{curse of dimensionality} \citep{Bellman_1957}, in which the training data do not sample all possible combinations of values in the $N$-dimensional feature space.
Furthermore, we normalise all of our features to better equip the machine to learn the significance of how given quantities relate to one another. For example, the value assigned to the growth in mass of a subhalo between two snapshots is more meaningful -- and aids physical interpretation -- if it is normalised by the total mass of the subhalo. The full list of features we include is given below:

\begin{itemize}
    \item \textbf{Subhalo mass}, $M_{\rm{subhalo}}$, normalised by $10^{12}\ \rm{M}_{\odot}$, the approximate subhalo mass at which galaxy formation is most efficient\footnote{This choice of normalisation is arbitrary and, since it is a constant value, does not affect the ML outputs. We opt to make the subhalo mass dimensionless for consistency with the other features.} \citep{behroozi_universemachine_2019}.
    \item $v_{\rm{max}}/v_{\rm{max, peak}}$: the ratio of a subhalo's \textbf{maximum rotational speed} at the snapshot in question to that of the greatest rotational speed it has had throughout its history. This is a measure of how much the subhalo has been stripped.
    \item \textbf{Mass concentration}, $M_{10\ \rm{pkpc}}/M_{\rm{subhalo}}$: the ratio of the mass in an aperture of radius $10\ \rm{pkpc}$, centred on the subhalo's centre of potential, to the total mass of the subhalo. Apertures of $10\ \rm{pkpc}$ will typically contain the majority of a subhalo's stellar component.
    \item \textbf{Dimensionless spin parameter}, $\lambda = \frac{J \sqrt{|E|}}{GM^{5/2}}$ (hereafter, referred to simply as the spin), as defined by \citet{Bullock_2001}. $J$ is the magnitude of the sum of the angular momenta of each of the subhalo's particles, relative to the subhalo's centre of potential; $E$ is the total energy of the subhalo; $M$ is its mass; $G$ is the gravitational constant.
    \item \textbf{Age}, $t_{1/2}/t_{\rm{snap}}$: the ratio of the age of the Universe, $t_{1/2}$, at which half the subhalo's mass had formed (relative to its mass at the root snapshot, $z=5$ or $z=10$) to the age of the Universe in the root snapshot, $t_{\rm{snap}}$.
    \item \textbf{Mass growth}, $\Delta M_{\rm{subhalo}}/M_{\rm{subhalo}}$: the difference in subhalo mass between the root snapshot and an earlier snapshot ($z=6$ for $z=5$ subhaloes and $z=12$ for $z=10$ subhaloes), normalised by the subhalo mass at the root snapshot. We choose the $z=6$ and $z=12$ snapshots for calculating $\Delta M_{\rm{subhalo}}$ as these are the snapshots closest in lookback time to $100\ \rm{Myr}$ (the approximate dynamical time of a galaxy) from the $z=5$ and $z=10$ snapshots, at $242\ \rm{Myr}$ and $105\ \rm{Myr}$, respectively\footnote{As we will discuss in Section \ref{sec:feat_imp}, we also ran a fiducial-resolution DMO simulation of Region 00 saved at snapshots $100\ \rm{Myr}$ apart to investigate the impact on model performance of using a higher snapshot cadence.}.
    \item $R_{1/2}/R_{\rm{vir}}$: the ratio of the subhalo's \textbf{half-mass radius}, $R_{1/2}$, to its approximate virial radius, $R_{\rm{vir}}$, where we define $R_{\rm{vir}}$ as $R_{\rm{vir}} = (3M_{\rm{subhalo}}/4\pi \times 200\rho_{\rm{crit}})^{1/3}$, where $\rho_{\rm{crit}}$ is the critical density at the redshift in question.
    \item A flag denoting whether a subhalo is a \textbf{central or a satellite}.
    \item $d_{\rm{central}}/R_{\rm{vir,\;central}}$: the \textbf{distance to the nearest central subhalo}, normalised by the virial radius of the nearest central subhalo. (This value is zero if the subhalo is itself a central.)
    \item \textbf{Redshift}. (5 or 10.)
\end{itemize}

We explored the inclusion of an overdensity feature, smoothed over a radius of $14/h\ \rm{cMpc}$, but found that it did not make a positive impact on model performance. \citet{thomas_first_2023} showed that encoding density information on multiple scales, ranging from $2.67\ \rm{cMpc}$ to $20\ \rm{cMpc}$, provides very little extra information about baryonic properties in \textsc{Flares} compared with a single overdensity measurement, so we chose not to explore other scales. \citet{lovell_machine_2021} and \citet{desanti_mimicking_2022} also found that the local density field makes little to no improvement in models trained to predict galaxy properties in simulations. More sophisticated measures of the large-scale environment, such as the skew of the subhalo mass distribution \citep{chittenden_2023}, will be explored in future work.

The $v_{\rm{max, peak}}$, $t_{1/2}$, and $\Delta M$ features required the construction of the DMO subhaloes' merger histories. We did this using the MErger Graph Algorithm (\textsc{Mega}) \citep{roper_mega_2020}. Each instance of the assembly history features was derived by tracing the main branch\footnote{The main branch is constructed by stepping backwards through time to the most massive progenitor at each timestep.} of the root subhalo's merger graph. All other features consist of values recorded at the $z=5$ and $z=10$ snapshots.

\subsection{Data pre-processing}
\label{section:pre-processing}

\subsubsection{Log transformation}

The ERT machine learns by minimising a loss function -- a measure of how close the predictions are to the true values. We use the mean squared error (MSE) as our loss function:
\begin{equation}
    \label{eqn:mse}
    \rm{MSE} = \frac{\sum_{\mathit{i}}^{\mathit{N}}\mathit{w_i}\left(y_{\rm{true},\mathit{i}} - y_{\rm{pred},\mathit{i}}\right)^2}{\mathit{N}}
\end{equation}
Here, $y_{\rm{true},\mathit{i}}$ is the 'true' value of instance $i$, as found in the simulations, and $y_{\rm{pred},\mathit{i}}$ is the machine-predicted value. $w_i$ is a weight, which we discuss below. The sum is performed over all $N$ instances in the training set. We note the importance of log-transforming the target variables prior to model training. Doing so reduces their dynamic range, and thus reduces the scale-dependence of fractional errors in the computation of the loss function \citep{jo_machine-assisted_2019}.
The vast majority of the galaxies included in our analysis are undergoing some level of star formation. In the few subhaloes that are not matched to star-forming galaxies, we assign an SFR equal to the lowest non-zero SFR permitted by the resolution: this allows us to log-transform all our target values.

\subsubsection{Bias correction}

Each instance in the training set contributes to the calculation of the loss function. If not adjusted for, the machine therefore seeks to minimise loss in the areas of parameter space that contain the bulk of the data. This is an issue when applying machine learning algorithms to imbalanced datasets where it may be important to accurately predict rare instances. Various methods to deal with this exist, including oversampling or undersampling the data (e.g. \citealt{desanti_mimicking_2022}). Here, we use the method of applying weights, $w$, to our training data. When computing the loss function, the model prioritises the correct prediction of samples with higher weights. We bin data in the training set according to subhalo mass and redshift, and apply a weight, $w_i$, to each instance, $i$, of $w_i = 1/N(i)$, where $N(i)$ is the total number of instances in the bin occupied by instance $i$. Thus, rare samples -- namely high-mass subhaloes and those at high redshift ($z=10$) -- are prioritised.

\subsubsection{Target variable normalisation}

We attempt to reduce the models' reliance on the most dominant feature: the subhalo mass. Stellar mass, SFR and metallicity are all tightly correlated with the mass of their subhalo. Such strong relationships can dominate the predictive reasoning of ML models, at the expense of potentially useful information residing in other input features. For example, \citet{lovell_machine_2021} found that their multi-feature ERT model for predicting stellar mass scarcely performed better than a model that used $v_{\rm{max}}$ -- a proxy for subhalo mass -- as a single input.
We normalise all instances of the log-transformed target variable, $f$, by its expected (mean) value at subhalo mass $x$, such that the new target value, $f'(x) = f(x) - E[f(x)]$\footnote{For each galaxy property, we obtain the mean as a function of subhalo mass by fitting cubic splines to the mean in bins of subhalo mass.}, is not correlated with subhalo mass. The target values are, therefore, a measure of the scatter above or below the mean in logspace. As we discuss in Appendix \ref{sec:appendix_variations}, this pre-processing step makes only minimal improvements to the models' outputs. However, its inclusion allows us to more easily explore the galaxy assembly bias -- that is, which subhalo properties, secondary to mass, are influential in determining the nature of their galaxies.

\subsubsection{Quality cuts}

We employ a mass cut of $M_{\rm{subhalo}} \geq 10^{10.5}\ \rm{M}_{\odot}$ to the DMO subhaloes included in our analysis, and include only those whose hydrodynamical matches have a summed mass within $\pm\ 0.5$ dex of their own.

\subsection{Model training and evaluation}
\label{sec:evaluation}

The best model hyperparameters\footnote{The hyperparameters we vary are \texttt{n\_estimators}, \texttt{max\_depth}, \texttt{min\_samples\_split}, \texttt{min\_samples\_leaf} and \texttt{max\_features}. All other hyperparameters are left to their default values, which can be found in the \textsc{Scikit-learn} documentation.} have been found using 5-fold cross validation with a 60:15:25 split in the percentage of data set aside for training, validation, and testing, respectively. The accuracy of the model in each case is determined using the $R^2$ metric (also known as the coefficient of determination):
\begin{equation}
\label{eqn:r2}
    R^2 = 1 - \frac{\sum_i (y_{\rm{true},\mathit{i}} - y_{\rm{pred},\mathit{i}}(\mathbf{\rm{x}}_\mathit{i}))^2}{\sum_i (y_{\rm{true},\mathit{i}} - \overline{y}_{\rm{true}})^2}
\end{equation}

We report our models' performances on the test set only, using $R^2$, the root mean squared error (RMSE; the square root of Equation \ref{eqn:mse}, sans weights), and Pearson's correlation coefficient:
\begin{equation}
\label{eqn:pearson}
    \rho = \frac{\rm{cov}\left(\mathit{y}_{\rm{true}},\ \mathit{y}_{\rm{pred}}\right)}{\sigma_{y_{\rm{true}}}\sigma_{y_{\rm{pred}}}}
\end{equation}

We seek to maximise $R^2$ and $\rho$, and minimise the RMSE. There are benefits to using multiple metrics. The RMSE has the same units as the quantity under study, so an RMSE of $x$ in logspace translates to an average error of $x$ dex, aiding interpretability. The advantage of $R^2$, conversely, is that its values are scale-independent. For a given ML pipeline, the target variable whose model has the highest $R^2$ score is the easiest variable to predict.
We also quote Pearson's correlation coefficient, $\rho$, as this is commonly used elsewhere in the literature and provides an at-a-glance insight into the tightness of the correlation between the predicted and ground truth values.

\section{Results at fiducial resolution}
\label{sec:results_eagle-res}

In this section, we present the results of our models trained and tested on the fiducial-resolution ($M_{\rm{DM}} = 1.15\ \times\ 10^7\ \rm{M}_\odot$) DMO simulations, incorporating all 40 \textsc{Flares} regions. We explore how well our models are able to replicate individual instances of the galaxy population -- and which features are the most important driving forces behind these predictions. We then examine how well we are able to replicate the simulated galaxy distributions by augmenting our ERT predictions with scatter intrinsic to the hydrodynamical simulations.

\subsection{ERT predictions}
\label{sec:predictions}

\begin{figure*}
 	\includegraphics[width=0.9\linewidth]{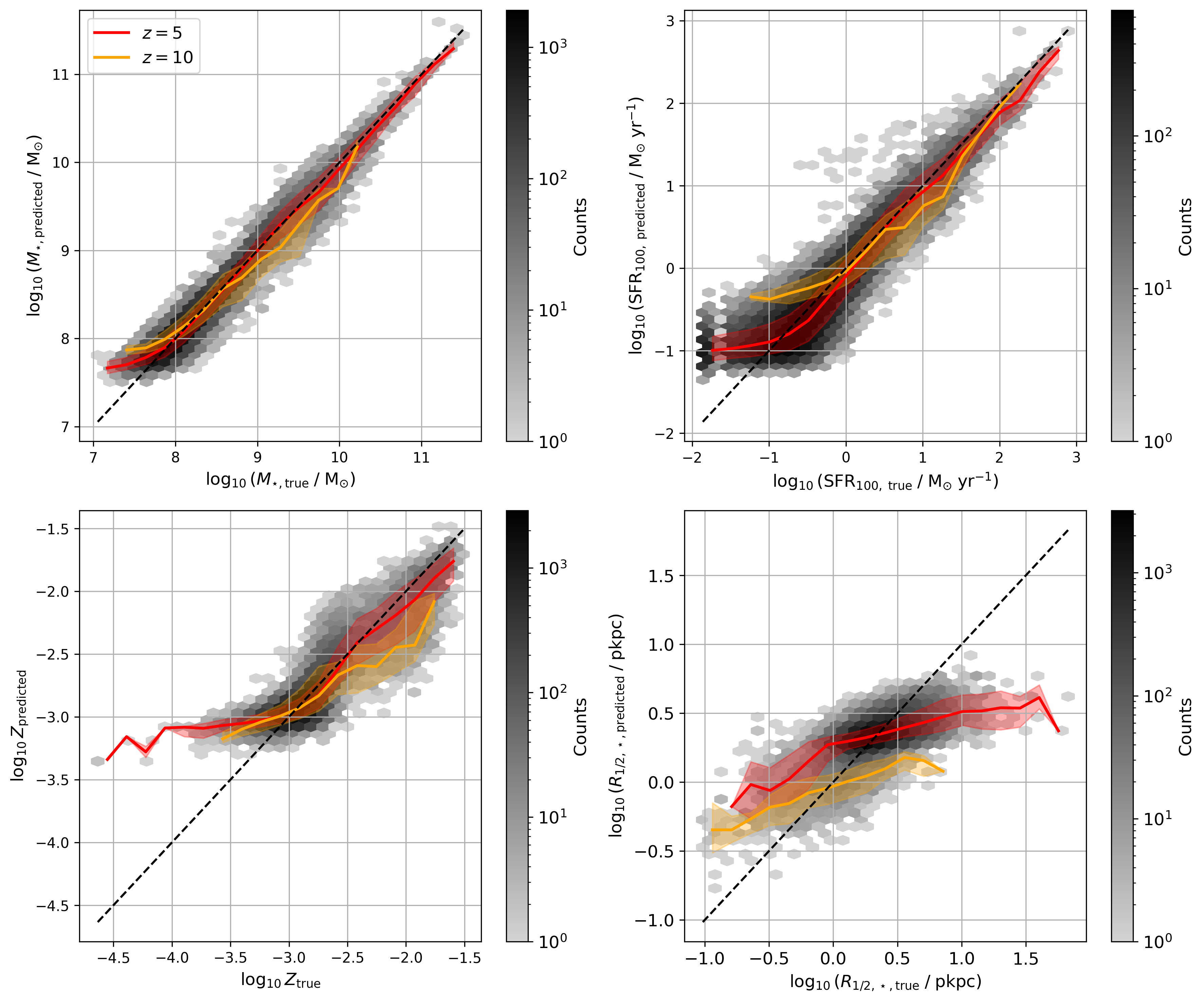}
	\caption{Predicted vs 'true' stellar mass, SFR, metallicity, and stellar half-mass radius, shown for all 40 \textsc{Flares} regions at fiducial resolution. We show the entire dataset, not just the test set. The hexbins show the results for the $z=5$ and $z=10$ data combined. The two coloured lines show the median predicted values in bins of true values, separately for $z=5$ and $10$, while the shaded regions around the lines show the 16th-84th percentile range of predicted values. The 'true' values are those from the hydrodynamical simulations, matched to the DMO-simulation subhaloes using the prescription outlined in Section \ref{section:halo_matching}.}
	\label{fig:preds_vs_true}
\end{figure*}

Figure \ref{fig:preds_vs_true} shows the predictions from our ERT models versus the ground truth (simulated) values in the test set for each galaxy property. The hexbins combine results from  $z=5$ and $z=10$, while the solid lines show the median prediction as a function of the ground truth for each redshift separately. Test-set performance metrics are reported in Table \ref{table:metrics}. Our results are comparable to -- and, in some cases, significantly improve upon -- previous attempts to machine learn the galaxy-halo connection (\citealt{kamdar_machine_2016_hydro}; \citealt{agarwal_painting_2018}; \citealt{lovell_machine_2021}; \citealt{Hausen_2023}; \citealt{hernandez_not_2023}; \citealt{das2024populatinggalaxieshalosmachine}).We note, however, that most of these studies focused on the low-redshift regime ($z \leq 2$).

We can see from Figure \ref{fig:preds_vs_true} that the distribution of predicted vs true values for stellar mass is tightly clustered around the one-to-one line, indicating strong model performance. This is reflected in the metrics in Table \ref{table:metrics}, which are strongest for stellar mass. Similar studies have been consistent in finding stellar mass the easiest galaxy property to predict (\citealt{agarwal_painting_2018}; \citealt{desanti_mimicking_2022}; \citealt{hernandez_not_2023}), owing to the nature of its cumulative growth and close connection to the parent halo mass.

Like stellar mass, the SFR predictions are mostly scattered symmetrically about the one-to-one line, exhibiting minimal bias at all but the lowest true values. The greater spread, however, reflects the more challenging nature of predicting the SFR. This is due to the SFR's sensitivity to local phenomena, such as gas accretion and tidal interactions, as well as feedback processes that are not encoded in the dark matter features. By predicting the SFR averaged over $100\ \rm{Myr}$, we reduce some of the difficulties that arise from short-term fluctuations due to processes such as gas inflows and feedback events \citep{Velazquez_2021}.

The predictions for metallicity and stellar HMR display larger biases; that is, true values at the low (high) end of the distribution are typically over(under)-predicted. Bias is a natural behaviour of deterministic ML algorithms. When no informative features are available from which to make more accurate predictions, the model will resort to predicting the median of the true distribution. The stronger biases in the metallicity and stellar HMR predictions, compared with stellar mass and SFR, suggests that dark matter halo features are worse equipped to predict these properties. Stellar HMR is the hardest property to predict (as shown quantitatively by its low $R^2$ score in Table \ref{table:metrics}). It is only weakly coupled to halo properties at high redshifts, while being sensitive to baryonic processes that occur only in the hydrodynamical simulations.

\begin{table}
\begin{tabular}{lccc}
                       & $R^2$ & Pearson's $\rho$ & RMSE  \\ \hline
$M_{\star}$       & $0.946 \pm 0.001$ & $0.973 \pm 0.001$      & $0.133 \pm 0.001$ \\
$\rm{SFR}_{100}$       & $0.814 \pm 0.004$ & $0.903 \pm 0.002$      & $0.310 \pm 0.003$ \\
$Z$                    & $0.739 \pm 0.007$ & $0.860 \pm 0.004$      & $0.140 \pm 0.002$ \\
$R_{1/2,\;\rm{stars}}$ & $0.295 \pm 0.010$ & $0.543 \pm 0.010$      & $0.170 \pm 0.002$
\end{tabular}
\caption{Model performances on the held-out test sets for each galaxy property studied. We quote the mean and standard deviation of the $R^2$, Pearson's correlation coefficient, $\rho$, and RMSE, across 1,000 bootstrappings of the test set.}
\label{table:metrics}
\end{table}

Central and satellite galaxies are known to display different characteristics \citep{Wang_2020}. We tested our models (trained on both central and satellite galaxy data) on the two types separately, and found stellar mass to be easier to predict in centrals than satellites, in line with expectations. Curiously, we obtained a higher $R^2$ score for stellar HMR in satellites compared with centrals, the statistical significance of which we validated with a two-sample z-test. We would expect the size of satellites to be harder to predict due to environmental effects such as tidal interactions. It is possible that the smaller sample of satellites in our dataset (just 9 per cent of the subhaloes are satellites) results in fewer edge cases. We found no statistically-significant difference in our models' ability to predict SFR and metallicity in centrals compared with satellites. We also looked at training on centrals and satellites separately and found no meaningful differences in model performance.
We performed the same comparative analysis on redshift, and found stellar mass, SFR, and metallicity to be easier to predict at $z=5$ than $z=10$. This can be attributed to fewer mergers \citep{Qiao_2024} and less bursty star formation \citep{Wilkins_2023}\footnote{To rule out the possibility that the better model performance at $z=5$ could be due to the larger amount of data available for the machine to learn from at this redshift, we re-trained our models on subsampled portions of the $z=5$ data. In each case, we still found stellar mass, SFR, and metallicity to be easier to predict at $z=5$ than $z=10$, so we conclude that the differences are physical.}. No significant difference was found for stellar HMR. Models tested on $z=10$ galaxies performed better when trained on both redshifts combined compared with when trained on $z=10$ data only, likely owing to the vast increase in data from which the machine was able to learn. In future endeavours, we will test the efficacy of building separate models for each redshift.

We found that our pre-processing steps of normalising the target variables and including sample weights in the model training made very little difference to the global performance metrics. Comparisons with results in which we remove these pre-processing steps are provided in Appendix \ref{sec:appendix_variations}.

\subsubsection{Feature importance}
\label{sec:feat_imp}

In this section, we explore the importance of each feature in determining the models' predictions.

\begin{figure*}
 	\includegraphics[width=0.9\linewidth]{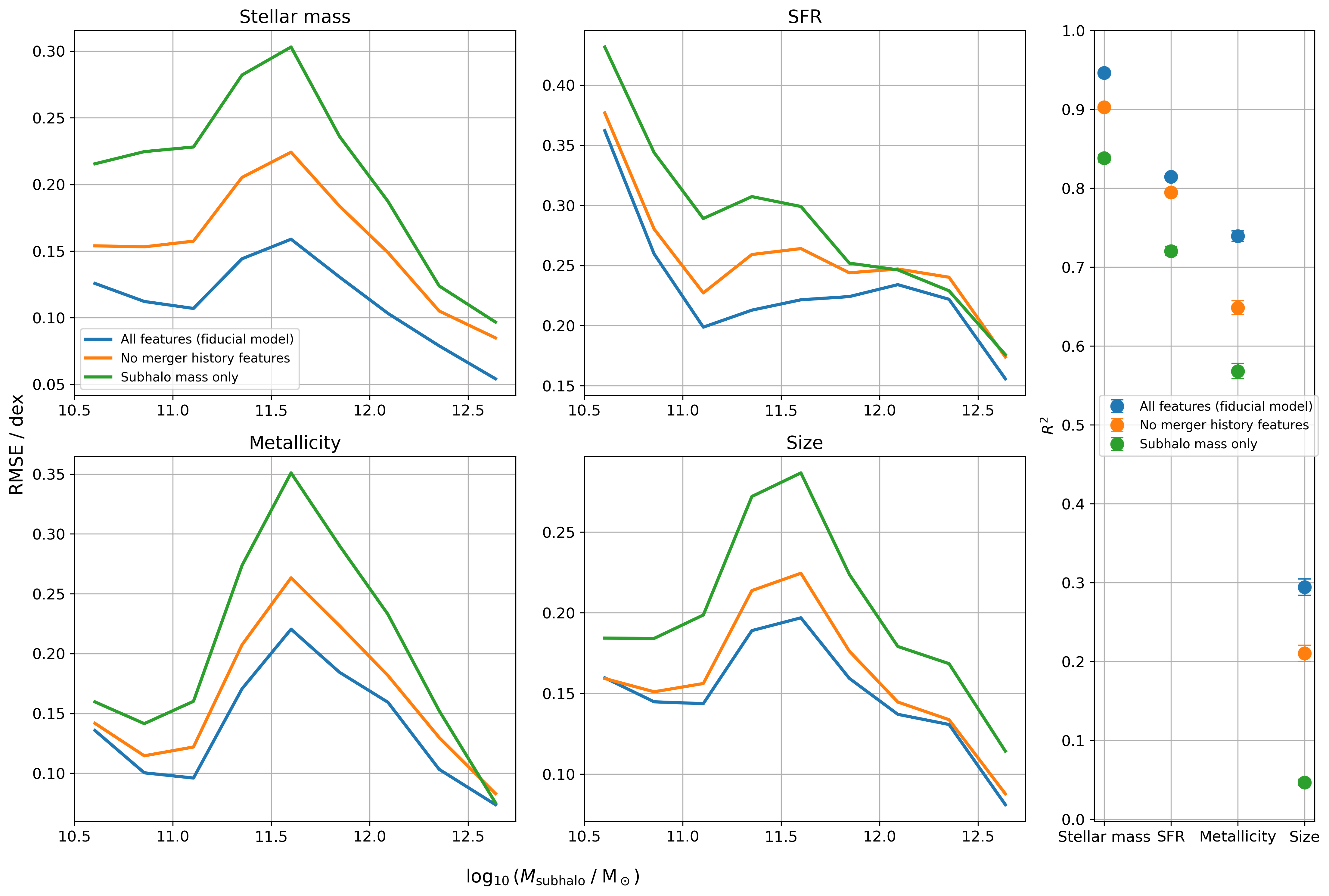}
	\caption{The RMSE between true and predicted values in bins of subhalo mass (four leftmost plots), and the overall $R^2$ scores (right-hand panel), for three different sets of models: our fiducial models (blue), one with merger history features removed (orange), and one trained on subhalo mass only (green).}
	\label{fig:metrics_features_included}
\end{figure*}

We begin by comparing our fiducial models with two simpler sets of models that use fewer features: one that uses subhalo mass only\footnote{For this model, we log the subhalo masses, which improves model performance. We do not weight the training data or normalise by $E[f(x)]$.} and one that includes all features listed in Section \ref{sec:features} except those obtained through the generation of merger graphs\footnote{We include $v_{\rm{max}}$ as a feature, but remove its normalisation by $v_{\rm{max,peak}}$.}. The former allows us to compare our fiducial models with an approach similar to a simple SHAM model, while the latter allows us to gain insight into the need -- or lack thereof -- for the tracing of merger histories in the $N$-body boxes to which we apply our models. We plot the RMSEs for each of these models as a function of subhalo mass in the leftmost panels of Figure \ref{fig:metrics_features_included}, while the right-hand panel shows the global $R^2$ scores.

This figure shows that, despite the strong connection between stellar and subhalo masses, our fiducial models vastly outperform a traditional SHAM approach, approximately halving the logspace error in stellar mass predictions at $M_{\rm{subhalo}} \sim 10^{11.5}\ \rm{M}_\odot$.
The inclusion of merger history features further improves predictions, particularly where the errors peak, at $M_{\rm{subhalo}} \sim 10^{11.5}\ \rm{M}_\odot$, though the improvement at the lowest and highest subhalo masses is minimal.
Rather surprisingly, the improvements made by including merger history features are smallest for SFR. We explored whether the coarse time resolution between lower-redshift snapshots (there is $\sim 242\ \rm{Myr}$ between the $z=5$ and $z=6$ snapshots) was responsible for this by re-running a DMO simulation of Region 00, saving snapshots at every $100\ \rm{Myr}$ interval (starting from $z=15$), in addition to at every integer redshift. For the mass growth feature, we then compared subhalo masses at $z=5$ with their progenitor masses at $z=5.385$, a time lapse of $104\ \rm{Myr}$. Despite the SFR being slightly more correlated with the mass growth between $z=5.385$ and $z=5$ compared with the mass growth between $z=6$ and $z=5$, we found no statistically-significant difference in model performance. Indeed, for each of our other target variables, increasing the time cadence did not improve the model predictions.

Having highlighted the importance of features beyond subhalo mass, we now assess the importance of each individual feature using SHapley Additive exPlanation (SHAP) values \citep{Lundberg-SHAP}, an approach derived from game theory. Shapley values measure how much each instance of each feature contributes to a model's prediction by considering all possible feature combinations. For a model with $m$ features, there are $2^m$ such combinations, making the calculation of Shapley values infeasible for models with many features and instances. Instead, we use SHAP, an approximation method for calculating Shapley values designed for cases where the number of features is large.
Unlike other feature importance methods, such as permutation importance \citep{altmann_permutation_2010}, which only provide a single importance value per feature, SHAP values are obtained for every \textit{instance} of each feature, allowing one to examine the extent to which different feature values affect the model predictions.
For any given instance and feature, the SHAP value is a marker of how much the feature value pushes the predicted target value up or down: the more positive the SHAP value, the more that particular feature value pushes the prediction up, relative to the mean predicted value, and vice versa.

\begin{figure*}
 	\includegraphics[width=0.9\linewidth]{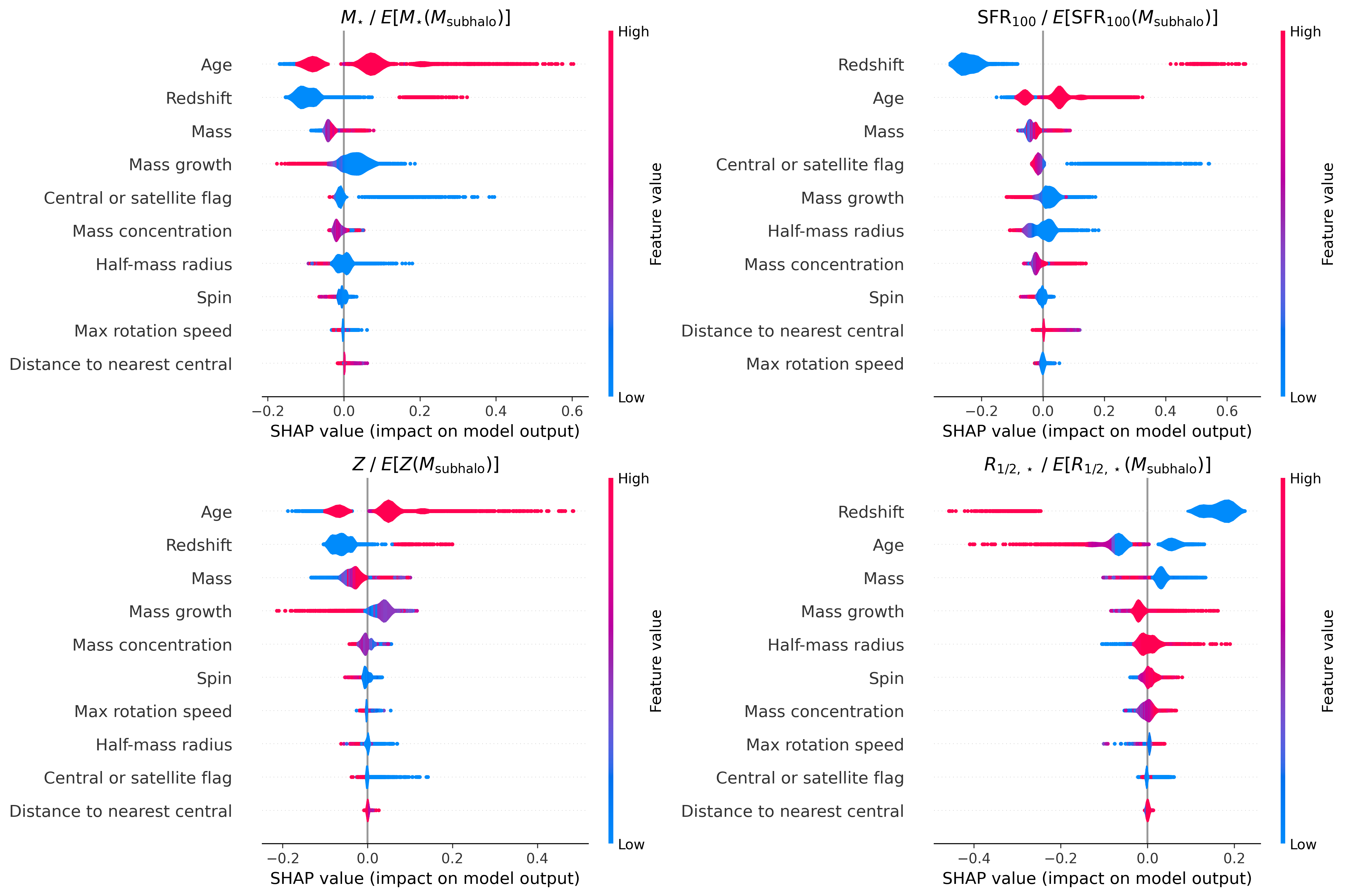}
	\caption{SHAP summary plots for each target variable. The target variable values, $f$, have been normalised by their expected values $E[f(x)]$ at subhalo mass $x$. For each galaxy property, each subhalo feature has a violin plot, showing the probability distribution of SHAP values amongst all instances of the test set. The subhalo features are listed in descending order from most to least important. For each feature, the violin plot is colour-coded according to how large, on average, the feature instances at the given SHAP value are, relative to other instances of the same feature.}
	\label{fig:shap_normalised}
\end{figure*}

We produce violin plots of the SHAP values for each of our target variables and features in Figure \ref{fig:shap_normalised}. The galaxy properties have each been normalised by their expected value as a function of subhalo mass (c.f. Section \ref{section:pre-processing}), so the plots illustrate the feature importances at \textit{fixed} subhalo mass\footnote{If this normalisation step is not taken, subhalo mass emerges as the most important feature for all target variables, except the stellar half-mass radius, for which mass concentration is the most important.}.

The features in each plot are ranked according to their overall importance to the model's outputs. For each target variable, each feature has a violin plot showing the probability distribution of the feature's SHAP values. The violin plots are colour-coded, such that high instances of the feature values are red, and low instances blue\footnote{For the central or satellite flag, centrals have value 1, and satellites 0, so centrals are coloured red and satellites blue.}. Redshift is one of the most important features in all cases, despite the distribution of redshifts in the dataset being highly imbalanced. (Given there are only two redshift values, $z=10$ datapoints are coloured red, and $z=5$ blue.) We can see from Figure \ref{fig:shap_normalised} that, at fixed subhalo mass, if an instance has redshift $z=10$, this increases the predicted value for stellar mass, SFR and metallicity, compared with if the redshift feature was absent. Meanwhile, $z=10$ values push predictions for the stellar half-mass radius down. This is consistent with our expectations: at fixed subhalo mass, high-redshift subhaloes are more compact and, therefore, form stars at a greater rate \citep{Xu2021}, enriching their environment in the process.
Age (defined as $t_{1/2}/t_{\rm{snap}}$) is also of very high importance, and follows the same pattern as redshift: higher instances increase predictions of stellar mass, SFR, and metallicity, while reducing predictions for the stellar HMR. Older galaxies have had more time to form stars and become enriched, but they have yet to become quenched.
Lower fractional mass growth also bolsters predictions for stellar mass, SFR and metallicity at fixed subhalo mass. This is likely because these galaxies are evolving in a more stable environment. Increasing mass growth is associated with tidal interactions, which suppress star formation and make galaxies more diffuse.
Instances in which the satellite flag is true are also associated with higher predictions for stellar mass, SFR, and metallicity, and lower predictions for stellar HMR. This concurs with the findings of \citet{roper_first_2023}, who showed that \textsc{Flares} galaxies with compact gas and stellar distributions are all satellites.

We note that, even when the target variables have been normalised by their mean value as a function of subhalo mass, the subhalo mass feature still improves model performance. Indeed, Figure \ref{fig:shap_normalised} shows that it remains one of the most important features in predicting each property. We surmise that this is because, for each target variable, the spread of scatter about the mean varies as a function of subhalo mass. Generally, the galaxy properties are well constrained by subhalo mass at the highest masses, but much less so at lower masses. Therefore, the value of $M_{\rm{subhalo}}$ still carries information that aids the machine: a decision node that splits according to subhalo mass provides information about the depth of the tree that's needed to arrive at an adequate answer.

To summarise, we find that at fixed subhalo mass, age and redshift are the most important informers of the galaxy properties, while the change in subhalo mass, and the subhalo's status as a central or satellite, are also important. Subhalo mass and mass concentration are highly important at variable subhalo masses, but less so when the subhalo mass is fixed.

\subsection{Augmenting the population statistics with intrinsic scatter}
\label{sec:adding_scatter}

We have shown that our ML models trained on subhalo features from DMO simulations do a good job at emulating the stellar masses, SFRs and metallicites of galaxies in the \textsc{Flares} simulations. However, the fact remains that some residuals are large, exceeding 1 dex. As we discussed in Section \ref{sec:intrinsic scatter}, perfection from models trained only on dark matter features is not possible due to the randomness introduced by the subgrid models in the hydrodynamical simulations. However, by modelling this randomness and incorporating it into our predictions post facto, we can hope to recover the simulated galaxy population statistics.

We add this scatter to our model predictions as follows.
\begin{itemize}
    \item For each DMO subhalo in regions 00 and 24, we have two realisations of the stellar mass, SFR, metallicity and stellar HMR, which come from two identical simulations run with different random number seeds; we denote these simulations 'Sim A' and 'Sim B'.
    \item For each galaxy property, at each redshift, we also compute the logspace RMSD between the two realisations as a function of subhalo mass (c.f. Figure \ref{fig:rmsd_rand_seed}).
    \item At each redshift, for each pair of galaxy properties, $p$ and $q$, we compute the correlation coefficients between the logspace differences in the Sim A and Sim B values of $p$ and $q$ as a function of subhalo mass. As we did for the computation of the RMSD (c.f. Section \ref{sec:intrinsic scatter}), we calculate the correlation coefficient for each instance based on the 10 nearest instances in subhalo mass. We then smooth out the noise by computing the average correlation coefficient in five equal-width bins in logspace. We fit splines\footnote{We fit cubic splines for $z=5$, and linear splines for $z=10$.} to the binned values, as shown in Figure \ref{fig:scatter_correlations} in Appendix \ref{sec:appendix_scatter_correlations}. Thus, for any subhalo of mass $M_{\rm{subhalo}}$, we have a measure of the typical random variation in each galaxy property, \textit{and} how this randomness correlates between each galaxy property\footnote{We expect that nearly all extractable information about the typical differences between two realisations of the galaxy population is found in the subhalo mass and redshift, so we avoid considering other subhalo properties.}. 
    \item For each subhalo in our ML datasets, we generate a vector of four random numbers drawn from Gaussian distributions with standard deviations of $\rm{RMSD}_{\mathit{i,j}}/\sqrt2$, where $\rm{RMSD}_{\mathit{i,j}}$ is the RMSD in property $j$ at subhalo mass $M_{\rm{subhalo},\mathit{i}}$\footnote{We normalise the RMSD by $\sqrt{2}$ since we are drawing from two Gaussian distributions and, therefore, adding in quadrature.}. We then apply a Cholesky decomposition \citep{benoit_1924} to the correlation matrix\footnote{Matrices must be positive-definite to perform a Cholesky decomposition. We encountered some instances in which numerical imprecision resulted in correlation matrices that were not strictly positive-definite. In these instances, we decomposed the eigenvalues of the matrix and reassigned any negative eigenvalues a small positive value ($10^{-8}$). We then reconstructed the correlation matrix using the adjusted eigenvalues.} at $M_{\rm{subhalo},\mathit{i}}$ and apply this transformation to the vector of random numbers.  
    \item Therefore, for each subhalo, we obtain a set of random numbers that are representative of the variance in each galaxy property in a subhalo of that mass, while preserving the correlation in the direction and magnitude of the variance in each galaxy property. We then add these random numbers to our model predictions. To achieve a more robust representation of the distributions obtained, we created 100 copies of each prediction and repeated this process 100 times.
\end{itemize}

We note that we are not the first to augment ML-predicted galaxy populations by adding scatter. \citet{das2024populatinggalaxieshalosmachine} added artificial scatter to their ML predictions by sampling from Gaussian distributions fine-tuned to replicate the ground truth. However, in this work, we have empirically determined the scatter that should be added. Furthermore, by measuring how the direction and magnitude of the scatter correlates between different galaxy properties, we ensure that the correlation that exists between galaxy properties in the simulations will be preserved in our mock catalogues.

\begin{figure*}
 	\includegraphics[width=\linewidth]{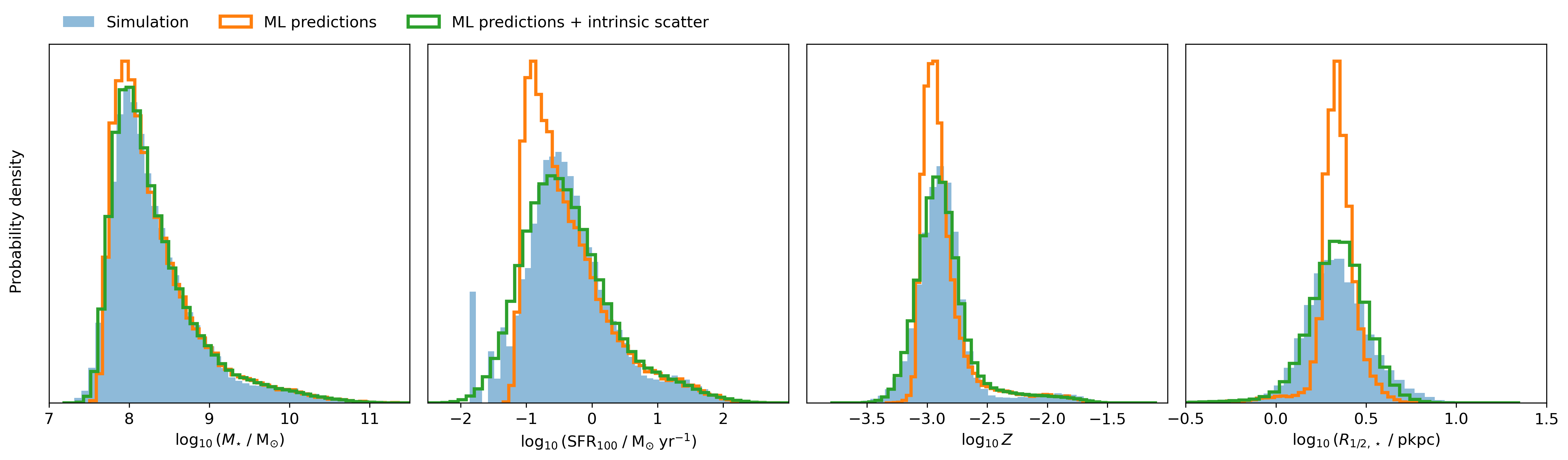}
	\caption{The probability distributions of the 'true' (blue, filled) and predicted (orange outline) galaxy properties in the \textsc{Flares} simulations, for $z=5$ and $z=10$ combined. The green line shows the distribution for our mock data, where the predicted galaxy properties have been augmented with intrinsic scatter. It can be seen that these distributions do a much better job of replicating the true distributions.}
	\label{fig:property_distributions}
\end{figure*}

We show the probability distributions of the galaxy properties in \textsc{Flares} in Figure \ref{fig:property_distributions}, along with the distributions that arise from our raw ERT predictions, and the ERT predictions augmented with intrinsic scatter.
The raw ML predictions alone do a good job of recovering the simulated stellar mass distribution. However, for each property, the peaks of the ML-predicted distributions are narrower than those of the simulated distributions -- particularly so for SFR, metallicity and stellar HMR. The predictions do a better job of replicating the tails of the distributions at the high end than they do at low values. This is because the larger values are more likely to be found in high-mass subhaloes, which we are able to model more accurately (c.f. Figure \ref{fig:metrics_features_included}). When we add intrinsic scatter to the predictions, the overall distributions are replicated very well. The addition of this scatter reduces the peaks of the predicted distributions and shifts them towards the median of the simulated distributions. It also extends the tails, particularly at lower values.

\textbf{\begin{figure*}
 \includegraphics[width=0.9\linewidth]{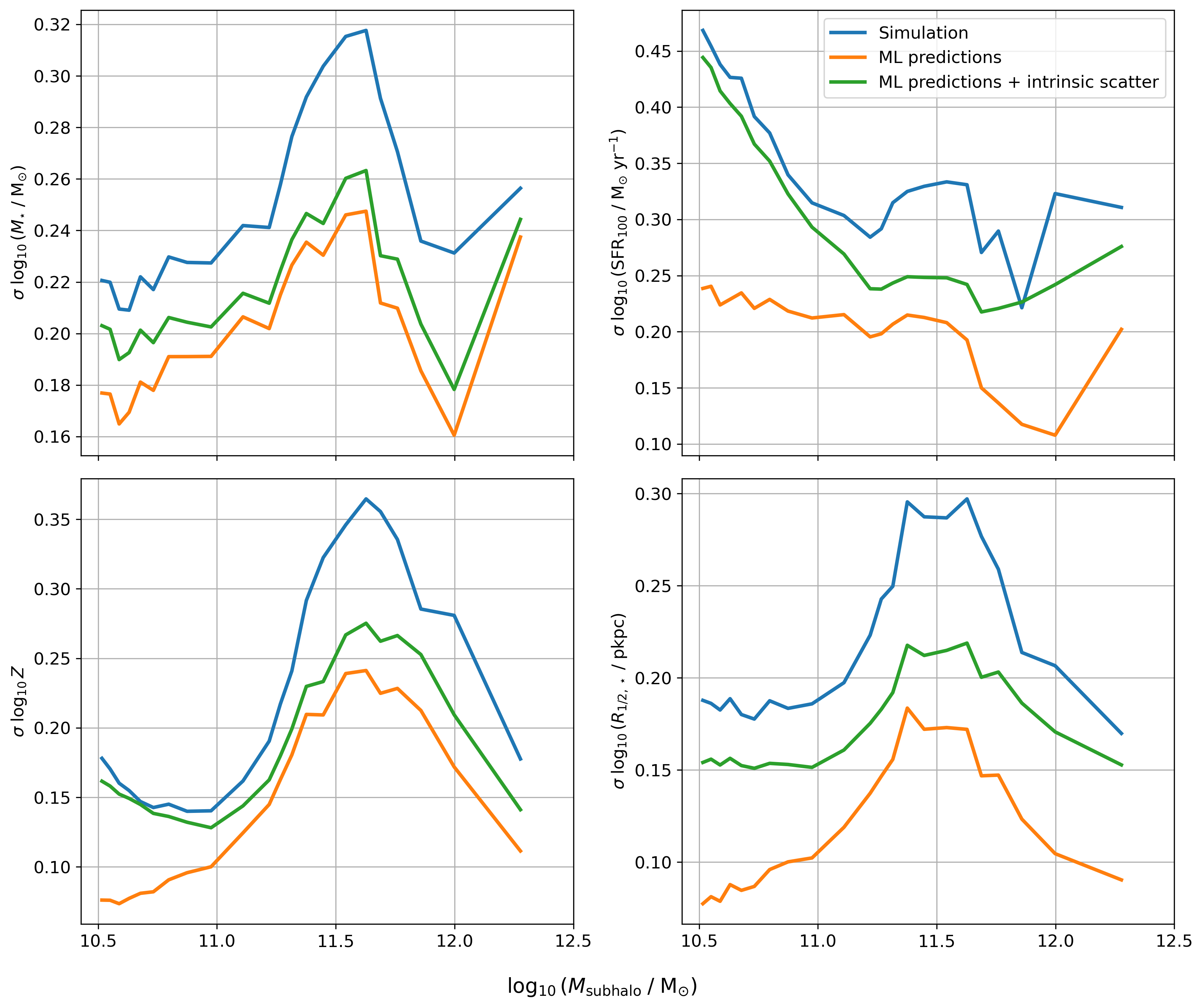}
	\caption{The standard deviation, $\sigma$, from the mean simulated (blue), predicted (orange) and mock (predicted with intrinsic scatter added; green) galaxy properties in bins of subhalo mass. The data include subhaloes from both $z=5$ and $z=10$.}
	\label{fig:scatter_vs_halo_mass}
\end{figure*}}

A key goal of our models is to replicate the spread in the values of galaxy properties at fixed subhalo mass -- something that simpler techniques, like SHAMs, typically neglect.
In Figure \ref{fig:scatter_vs_halo_mass}, we show the standard deviation ($\sigma$) in the simulated and predicted values of galaxy properties in bins of subhalo mass.

The scatter at fixed subhalo mass tends to peak around $M_{\rm{subhalo}} \sim 10^{11.5}\ \rm{M}_\odot$. This likely has the same origin as the bump observed at similar masses in Figure \ref{fig:rmsd_rand_seed}, namely AGN feedback. SFR is an exception -- the scatter in SFR peaks at the lowest masses, where star formation is more bursty.
It can be seen that the ML models fail to capture the full extent of the scatter. We improve upon this when we add intrinsic scatter to our predictions. However, despite the much-improved distributions obtained in Figure \ref{fig:property_distributions}, the full standard deviation as a function of subhalo mass is not reproduced, showing that there is still room for improvement.

We find that including the galaxy properties not under immediate analysis as additional input features in the ML models -- an approach first explored by \citet{desanti_mimicking_2022} -- \textit{does} lead to the standard deviations being fully replicated. We emphasise that including features from hydrodynamical simulations defeats the purpose of using machine learning to create mock catalogues, as no computational expense is spared. However, this result proves that the galaxy distributions can be entirely recovered by ML models, given the right set of input features. \citet{jo_machine-assisted_2019} found that they were able to improve their predictions for SFR by training an initial model to predict stellar magnitudes from halo properties, and using these stellar magnitude predictions as inputs in a second-stage model. To what extent we can better recover the distributions in \textsc{Flares} by refining our methods and choice of DMO-simulation features, as well as using predicted galaxy properties as inputs, will be the subject of further study.

\subsubsection{Universal distribution functions}
\label{sec:dist_funcs}

\begin{figure*}
    \includegraphics[width=0.9\linewidth]{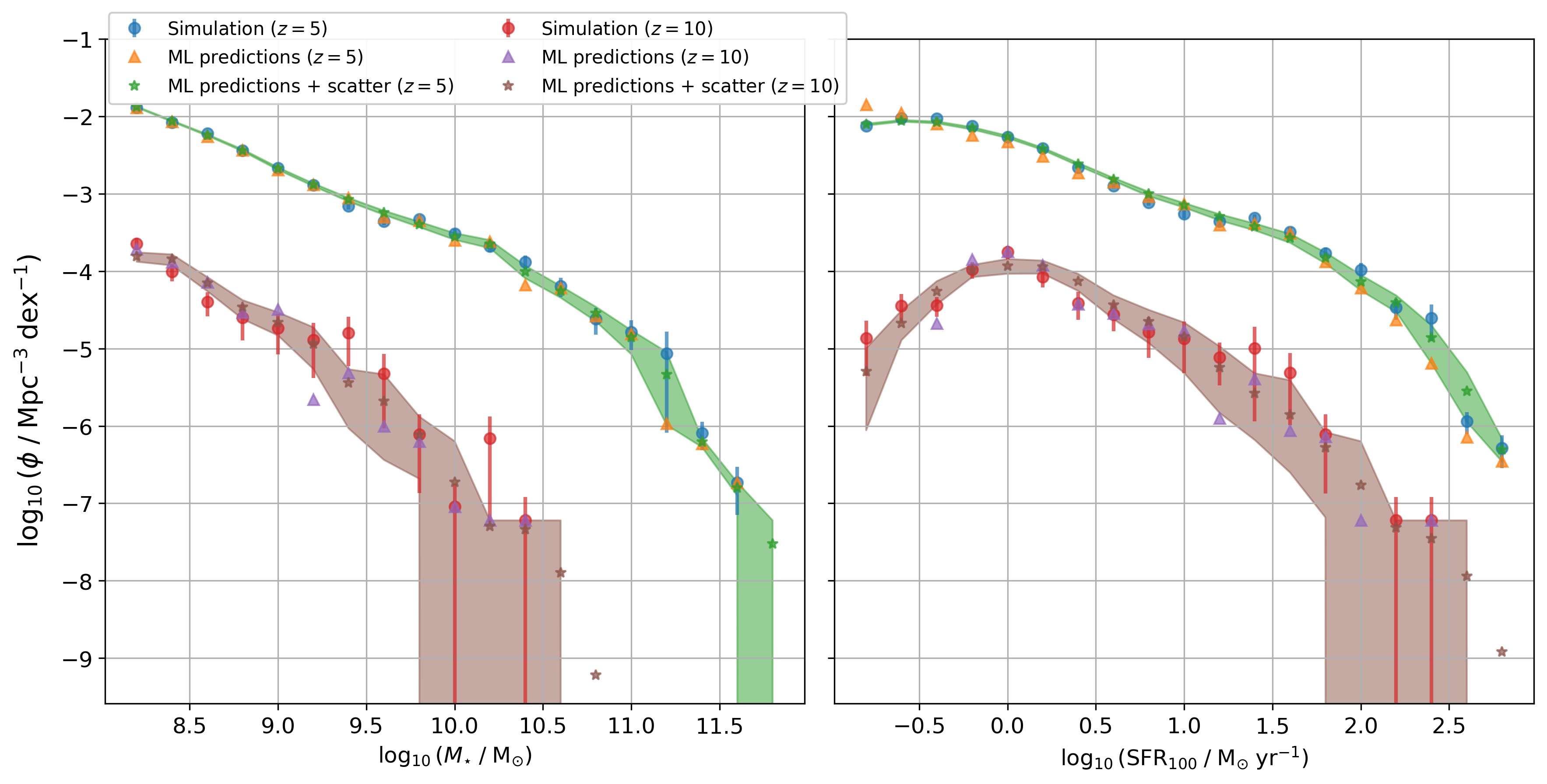}
    \caption{The galactic stellar mass function (GSMF; left) and star formation rate function (SFRF; right) for the $(3.2\ \rm{cGpc})^3$ volume, determined using the \textsc{Flares} weighting scheme. Functions are shown for the simulation, ML predictions, and mock values (ML predictions + intrinsic scatter) at $z=5$ and $z=10$. $1\sigma$ Poisson uncertainties are shown for the simulations. The mock values are averaged over 100 iterations; the shaded areas show the 16th-84th percentile range.}
    \label{fig:smf_sfrf}
\end{figure*}

We show the Universal galactic stellar mass function (GSMF) and star formation rate function (SFRF)\footnote{These functions have a form equivalent to Equation \ref{eqn:mass_function}, but with $M_{\rm{subhalo}}$ replaced by $M_{\star}$ and SFR, respectively.} in Figure \ref{fig:smf_sfrf} for the \textsc{Flares} simulations, our ML predictions, and our mock values (ML predictions enhanced with intrinsic scatter).
For each \textsc{Flares} region, $r$, we compute the number density of true, predicted, and mock stellar masses and SFRs in bins 0.2 dex in width. These number densities are weighted by the region's weight, $w_r$ (c.f. Section \ref{section:halo_matching}), and then summed. For the simulations, we show the $1\sigma$ Poisson uncertainties as error bars. For our mock values of $\phi$, we show the 16th-84th percentile range, obtained over 100 independent scatter generations, as shaded regions.

We see that the simulated GSMF is reproduced extremely well by the raw predictions alone at $z=5$. In nearly all bins, the ML-predicted value of $\phi$ is within the $1\sigma$ errors of the simulations. The raw ML predictions are outperformed by our scatter-augmented values, however, all of which lie within the $1\sigma$ errors of the simulations. The ML model finds it slightly harder to replicate the GSMF at $z=10$. The addition of scatter does little to change the predicted GSMF at this redshift but our mock values of $\phi$ are still within the $1\sigma$ errors of the simulation in $\sim70$ per cent of mass bins.
For the SFRF, our scatter-augmented predictions outperform the raw predictions at $z=5$. In nearly all bins, our mock values of $\phi$ lie within the simulated $1\sigma$ errors. At $z=10$, there are a few instances of the $\phi$ values from the raw ML predictions outperforming those from the scatter-augmented predictions. We can see from Figure \ref{fig:property_distributions} that the distributions of the ML predictions and scatter-augmented values are very similar at the stellar mass and SFR values at which these instances occur. The few instances in which $\phi$ is better for the raw ML predictions may point to the stochasticity in stellar mass and SFR possessing second-order dependences on the large-scale environment, which only become apparent after we apply weights to the regions. We will re-run more regions with different random number seeds to investigate this further.

\section{Results at lower mass resolutions}
\label{sec:results_low-res}

Having tested our models on the fiducial-resolution \textsc{Flares} simulations, we now assess their performances when applied to DMO simulations run at lower resolutions.  This will provide insight into which $N$-body boxes we can apply our models to.
As outlined in Section \ref{sec:dmo}, we have re-run DMO simulations of Region 00 at three lower resolutions. We label these resolutions $N/8$, $N/64$, and $N/512$. The number of subhaloes in each of these simulations, and the numbers included in our analysis, are outlined in Table \ref{table:low-res_halos}. At each resolution, we assign galaxy properties to the DMO subhaloes using the method outlined in Section \ref{section:halo_matching}\footnote{The fraction of DMO subhaloes with multiple hydrodynamical matches increases as the resolution decreases. Since we define the stellar mass and SFR in a DMO subhalo to be the sum contribution from their matching galaxies, we checked whether this approach would lead to any out-of-distribution errors. We found that the maximum stellar mass at the lowest resolution is only 0.01 dex larger than the greatest stellar mass found at fiducial resolution, while the SFR values at all resolutions are within the range found at fiducial resolution.}.

\begin{figure}
 	\includegraphics[width=\columnwidth]{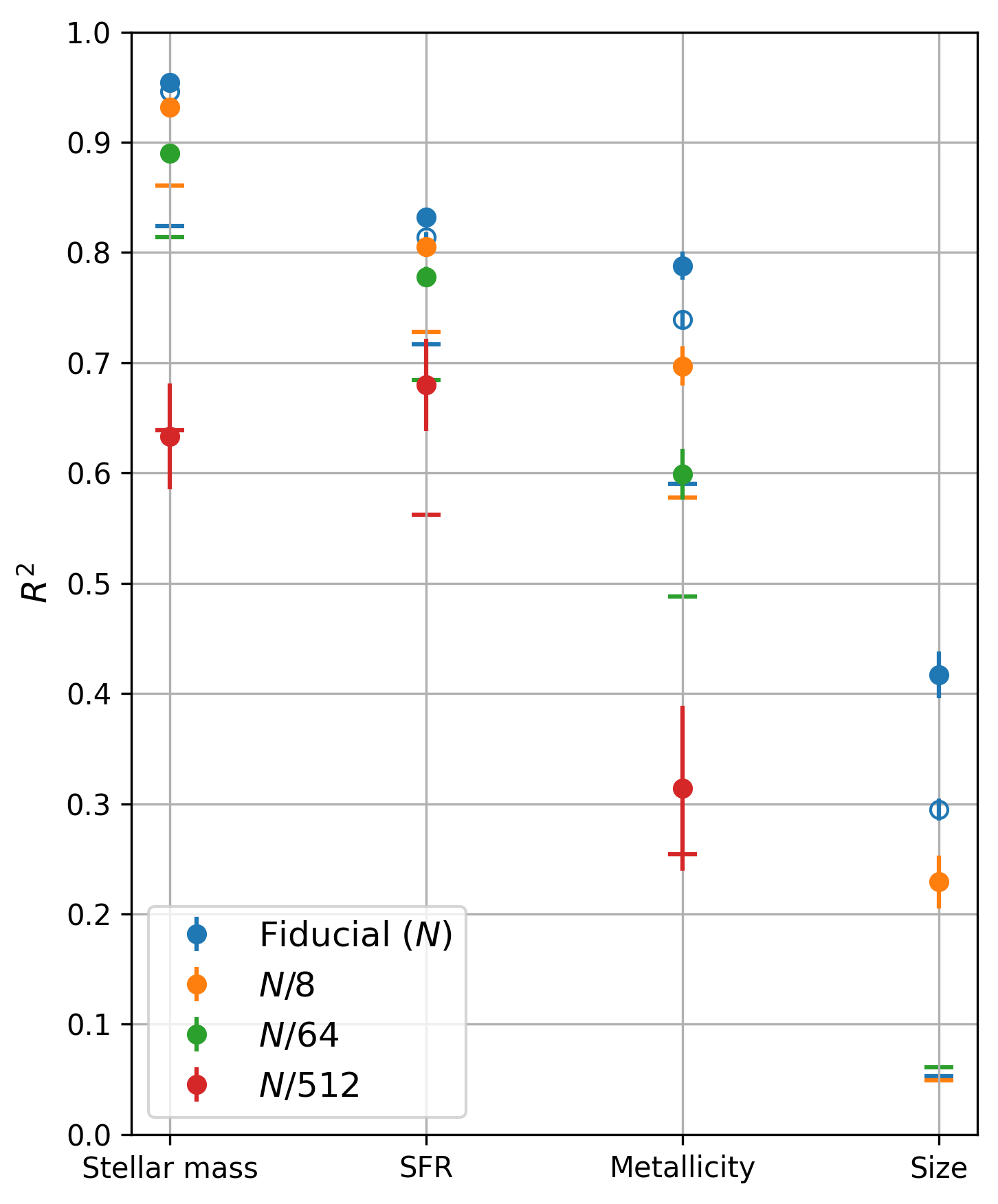}
	\caption{For each target variable, the solid circles show the $R^2$ scores for our fiducial models (described in Section \ref{sec:ml}) applied to four DMO runs of Region 00, differing only in resolution. The error bars show the standard deviation in the $R^2$ scores across 1,000 bootstrapped samples of the test set. For reference, we also show the $R^2$ scores for our fiducial models tested on all 40 fiducial-resolution regions (hollow blue circles; c.f. Section \ref{sec:results_eagle-res}), and the $R^2$ scores for models trained on subhalo mass only (horizontal lines).}
	\label{fig:mass_resolution_dependence}
\end{figure}

We plot the $R^2$ scores for each of our fiducial models applied to the lower-resolution DMO simulations in Figure \ref{fig:mass_resolution_dependence}. We also show the results of the same models tested on the fiducial-resolution run of Region 00 for comparison. For stellar mass and SFR, the $R^2$ score decreases by no more than $0.03$ and $0.06$ as we decrease the resolution to $N/8$ ($M_{\rm{DM}} = 8.92 \times 10^7\ \rm{M}_\odot$) and $N/64$ ($M_{\rm{DM}} = 7.14 \times 10^8\ \rm{M}_\odot$), respectively. This means that our fiducial models applied to DMO simulations of $N/8$ resolution outperform the models that lack merger history features when applied to EAGLE-resolution DMO simulations. Meanwhile, the performance of the latter is approximately matched by our fiducial models applied to an $N/64$-resolution simulation. This is a hugely positive result as it means we can apply our models to $N$-body boxes with mass resolutions approximately one order of magnitude lower than EAGLE, without a significant loss of precision. The state-of-the-art FLAMINGO simulations have been run with dark matter-only at a resolution of $M_{\rm{DM}} = 8.40 \times 10^8\ \rm{M}_\odot$ (slightly worse than our $N/64$ resolution) in a $(1.0\ \rm{cGpc})^3$-volume box \citep{flamingo_schaye_2023}, while the Illustris-TNG suite contains a DMO simulation of volume $(303\ \rm{cMpc})^3$ run at a resolution of $M_{\rm{DM}} = 4.73 \times 10^7\ \rm{M}_\odot$ \citep{springel_2018} -- \textit{better} than the resolution of our $N/8$ simulation. This goes to illustrate the size of the mock catalogues our models have the potential to generate.

Even for metallicity and stellar HMR, where degradation in model performance is more noticeable as the resolution decreases, our fiducial models perform better when applied to the $N/8$-resolution simulation than models lacking merger history features do when applied to the EAGLE-resolution simulations.
Stellar HMR, being the hardest property to predict, is the most hampered by resolution. At $N/64$ and $N/512$ ($M_{\rm{DM}} = 6.52 \times 10^9\ \rm{M}_\odot$)  resolutions, the $R^2$ scores for stellar HMR decrease below zero. In these cases, simple scaling relations would outperform our models. Indeed, at $N/512$ resolution, our fiducial models for stellar mass and metallicity also perform no better than models trained on subhalo mass only (as denoted by the horizontal lines). At this resolution, the $R^2$ scores for models trained on subhalo mass only are also shown to decrease sharply, further illustrating the decoupling between galaxy and halo properties at very low resolutions.

For each resolution, we also explored re-training the models on the DMO features at the resolution in question. We found this made minimal difference to the results at $N/8$ and $N/64$ resolution, suggesting that the models trained at fiducial resolution are able to generalise to the lower-resolution DMO data down to the $N/64$ level. At $N/512$ resolution, performances are slightly better for the re-trained models, indicating that -- at very low resolutions -- there are resolution-dependent differences in the DMO features that bias the performance of the fiducial models.

\section{Summary and conclusions}
\label{sec:conclusions}

In this study, we have trained deterministic ML models -- using the popular extremely randomised trees (ERT) algorithm \citep{geurts_extremely_2006} -- to predict four key galaxy properties (stellar mass, time-averaged SFR, metallicity, and stellar half-mass radius) in the First Light and Reionisation Epoch Simulations (\textsc{Flares}; \citealt{lovell_first_2020}; \citealt{vijayan_first_2020}) suite of 40 zoom simulations. Learning these relationships in a wide range of environments allows us to map galaxies onto haloes in large $N$-body boxes, enabling the efficient generation of large mock galaxy catalogues that can be used for comparisons with wide-field observational surveys.

By running additional versions of two of the hydrodynamical zoom simulations, differing only in their random number seed, we have quantified the scatter in the galaxy-halo relations that cannot be explained by dark matter properties alone. Modelling this scatter has enabled us to add it to our ML predictions in post-processing, resulting in better replications of the simulated galaxy distributions.

We discuss our main conclusions below.
\begin{itemize}
    \item Features beyond subhalo mass are of critical importance to our ML models. Perhaps surprisingly, this is most evident for stellar mass; the galaxy property most strongly correlated with subhalo mass. Compared with models that use subhalo mass as their only input feature, models that include additional features taken from the root snapshot reduce the logspace RMSE by $25$ per cent, $16$ per cent, $10$ per cent, and $9$ per cent for stellar mass, SFR, metallicity, and stellar half-mass radius, respectively. Models that also include merger history features reduce the logspace RMSEs by a further $30$ per cent, $5$ per cent, $15$ per cent, and $6$ per cent, respectively. These figures highlight the importance of tracing the assembly history when predicting stellar mass.
    \item Except for SFR, where bursty star formation suppresses model accuracy in the least massive subhaloes, the errors in our models peak in subhaloes of mass $\sim 10^{11.5}\ \rm{M}_\odot$. It's also at this mass that we observe larger differences between simulations run with different random number seeds (c.f. Figure \ref{fig:rmsd_rand_seed}) and an increase in scatter in the galaxy-halo relations (c.f. Figure \ref{fig:scatter_vs_halo_mass}). We can attribute the former to AGN feedback. However, $\sim 10^{11.5}\ \rm{M}_\odot$ is also the mass at which our model errors are most reduced by including features beyond subhalo mass (c.f. Figure \ref{fig:metrics_features_included}), suggesting that some of the scatter in the galaxy-halo relations at this mass can be explained by dark matter properties.
    \item At a fixed subhalo mass, the most important features in predicting galaxy properties are the snapshot redshift and the time since half the subhalo's mass had formed.
    \item Despite the success of the ML models (e.g. $R^2$ scores in excess of 0.94 for $M_\star$), they are not able to replicate the full distribution of galaxy properties. This is due to the stochasticity inherent to the subgrid models in the hydrodynamical simulations. Changes to the numerics at these unresolved scales induce chaos-like behaviour. We surmise -- due to the non-monotonic dependence on subhalo mass of the differences in like-for-like galaxies -- that these changes lead to real physical differences at resolved scales. Modelling this stochasticity, and using it to add realistic scatter to our ML predictions, leads to a much better reproduction of the simulated galaxy populations.
    \item We have explored how well our models -- trained at EAGLE resolution -- generalise to features from lower-resolution DMO simulations. The global statistics of ML-predicted galaxy populations have previously been shown to be insensitive to decreases in the DMO-simulation mass resolution of a factor of 8 (\citealt{lovell_machine_2021}; \citealt{deAndres_2023}). Here, we have shown that depreciation in our model predictions for \textit{individual} subhaloes is also minimal when reducing the DMO-simulation resolution by this amount, while our stellar mass and SFR models still perform well when applied to a simulation 64 times lower in resolution. This gives us confidence that we can use our models to generate mock catalogues using existing $N$-body boxes that span up to a Gpc in length, e.g. FLAMINGO \citep{flamingo_schaye_2023}.
\end{itemize}

Given the stochasticity in the simulated galaxy properties associated with a fixed set of subhalo properties, we will focus our future efforts on using probabilistic ML methods, such as Gaussian processes \citep{Rasmussen_2006}, diffusion models \citep{dickstein_2015}, or normalising flows (\citealt{papamakarios_2021}; \citealt{rodrigues_2024}). We will predict photometric properties for the \textsc{Flares} catalogue, obtained through the new synthetic observations package \textsc{Synthesizer}\footnote{https://github.com/flaresimulations/synthesizer/} (Lovell et al., in prep.), and apply our models to large \textit{N}-body boxes, thus making possible direct comparisons with observations.

\section*{Acknowledgements}
We thank the EAGLE team for developing the simulation code. We also wish to acknowledge the following open-source software packages used in our analysis: \textsc{Scikit-learn} \citep{scikit-learn}, SciPy \citep{2020SciPy-NMeth}, AstroPy \citep{astropy}, and Matplotlib \citep{Hunter:2007}.

This work used the DiRAC@Durham facility managed by the Institute for Computational Cosmology on behalf of the STFC DiRAC HPC Facility (www.dirac.ac.uk). The equipment was funded by BEIS capital funding via STFC capital grants ST/K00042X/1, ST/P002293/1, ST/R002371/1 and ST/S002502/1, Durham University and STFC operations grant ST/R000832/1. DiRAC is part of the National e-Infrastructure. The Eagle simulations were performed using the DiRAC-2 facility at Durham, managed by the ICC, and the PRACE facility Curie based in France at TGCC, CEA, Bruyeres-le-Chatel.

MGAM wishes to acknowledge the help and support of his late supervisor, Peter Thomas, without whom this paper would not have been written.

MGAM and LTCS acknowledge support from their STFC studentships. WJR acknowledges support from the Sussex Astronomy Centre STFC Consolidated Grant
(ST/X001040/1). 

We list here the roles and contributions of the authors according to the Contributor Roles Taxonomy (CRediT)\footnote{https://credit.niso.org/}. \textbf{Maxwell Maltz}: Conceptualization, Data Curation, Formal Analysis, Investigation, Methodology, Visualization, Writing - original draft. \textbf{Peter Thomas}: Conceptualization, Methodology, Supervision, Writing - review and editing. \textbf{Christopher Lovell, William Roper}: Conceptualization, Methodology, Writing - review and editing. \textbf{Aswin Vijayan}: Methodology, Writing - review and editing. \textbf{Dimitrios Irodotou, Shihong Liao, Louise Seeyave}: Writing - review and editing.

%%%%%%%%%%%%%%%%%%%%%%%%%%%%%%%%%%%%%%%%%%%%%%%%%%
\section*{Data Availability}
\textsc{Flares} data and analysis code are available at https://github.com/flaresimulations. The raw data and ML code underpinning this analysis will be made available upon request.



\bibliographystyle{mnras}
\bibliography{FLARES_ML_paper} % if your bibtex file is called example.bib

%\begin{thebibliography}{99}
%\end{thebibliography}

%%%%%%%%%%%%%%%%%%%%%%%%%%%%%%%%%%%%%%%%%%%%%%%%%%

%%%%%%%%%%%%%%%%% APPENDICES %%%%%%%%%%%%%%%%%%%%%

\appendix
\section{Subhaloes in the low-resolution simulations}
Table \ref{table:low-res_halos} shows the number - and percentage of successfully matched - DMO subhaloes, for each of our resolution-variant runs of Region 00. As the resolution decreases, the percentage of DMO subhaloes whose matches have combined masses within $\pm 0.5$ dex of their own mass also decreases. Up to $5.2$ per cent of subhaloes at the lowest resolution are discarded. However, since we only use this data for testing - not training - this does not introduce biases. Therefore, we maintain the same selection criteria across all resolutions to allow for direct comparison.
The number of DMO subhaloes with $M_{\rm{subhalo}} \geq 10^{10.5}\ \rm{M}_\odot$ increases as we go down to $N/64$ resolution due to fewer sub-components being resolved within these structures. The number of DMO subhaloes with $M_{\rm{subhalo}} \geq 10^{10.5}\ \rm{M}_\odot$ then decreases again at $N/512$ resolution as the minimum resolved mass exceeds the imposed mass cut.

\begin{table*}
\centering
\resizebox{\textwidth}{!}{
\begin{tabular}{lcc|cc|cc|cc}
                                                   & \multicolumn{2}{c|}{Fiducial resolution} & \multicolumn{2}{c|}{$N/8$ resolution} & \multicolumn{2}{c|}{$N/64$ resolution} & \multicolumn{2}{c}{$N/512$ resolution} \\ \cline{2-9} 
                                                   & $z=5$                & $z=10$            & $z=5$               & $z=10$          & $z=5$              & $z=10$            & $z=5$              & $z=10$            \\ \hline
DMO subhaloes with $M \geq 10^{10.5}\ \rm{M}_\odot$ & 2,166                & 53                & 2,423               & 58              & 2,672              & 61               & 310                & 2                 \\
... Matched to hydro subhaloes...                   & 2,157 (99.6\%)       & 52 (98.1\%)       & 2,407 (99.3\%)      & 57 (98.3\%)     & 2,646 (99.0\%)     & 61 (100.0\%)     & 309 (99.7\%)       & 2 (100.0\%)       \\
... Masses of matches within 0.5 dex...            & 2,146 (99.1\%)       & 52 (98.1\%)       & 2,380 (98.2\%)      & 56 (96.6\%)     & 2,567 (96.1\%)     & 59 (96.7\%)      & 294 (94.8\%)       & 2 (100.0\%)       \\
... Matches contain galaxies...                    & 2,146 (99.1\%)       & 52 (98.1\%)       & 2,380 (98.2\%)      & 56 (96.6\%)     & 2,567 (96.1\%)     & 59 (96.7\%)      & 294 (94.8\%)       & 2 (100.0\%)       \\
... Matches contain star formation...              & 2,121 (97.9\%)       & 52 (98.1\%)       & 2,349 (97.0\%)      & 56 (96.6\%)     & 2,499 (93.5\%)     & 59 (96.7\%)      & 294 (94.8\%)       & 2 (100.0\%)      
\end{tabular}}
\caption{The DMO subhalo statistics in Region 00, run at four different resolutions. For each simulation and snapshot, we display the number of subhaloes with $M_{\rm{subhalo}} \geq 10^{10.5}\ \rm{M}_{\odot}$; the number above this mass cut that match to at least one hydrodynamical subhalo; the number of these subhaloes whose masses are within 0.5 dex of the summed masses of their matches; the number of these subhaloes that have at least one match with a stellar component; and the number of these subhaloes that match to at least one star-forming subhalo. For each cell, the percentage in brackets denotes the percentage of subhaloes with $M \geq 10^{10.5}\ \rm{M}_{\odot}$ that fulfill the criteria up to and including that column. We discard no more than 5.2 per cent of subhaloes with $M_{\rm{subhalo}} \geq 10^{10.5}\ \rm{M}_{\odot}$ at any resolution and snapshot.}
\label{table:low-res_halos}
\end{table*}
\label{sec:appendix_low-res}
\section{Mass ratios of matching subhaloes}
\begin{figure}
    \includegraphics[width=0.5\textwidth]{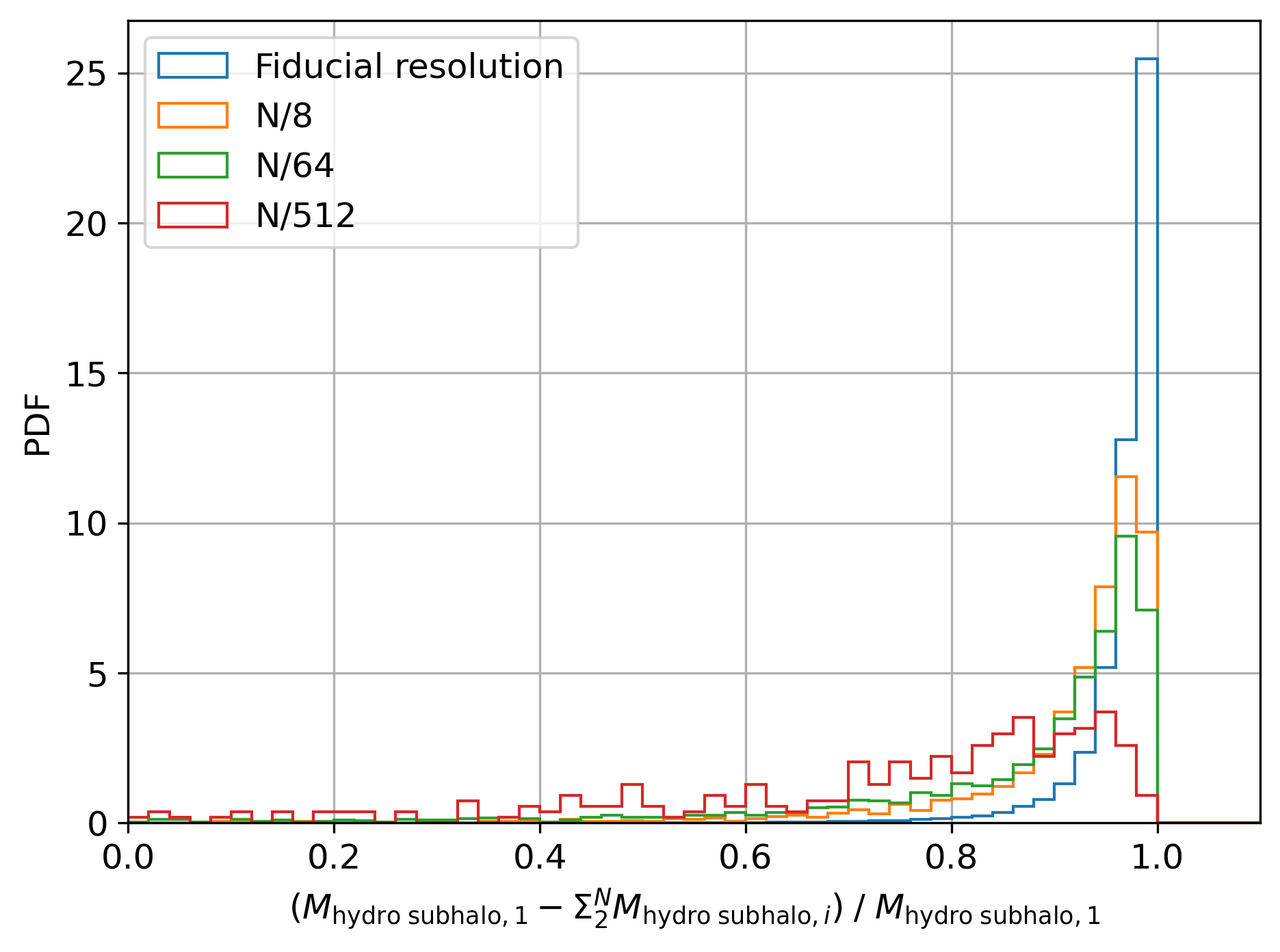}
	\caption{The probability distribution functions (PDFs) for the mass ratios of hydrodynamical subhaloes matched to DMO subhaloes. For each DMO-simulation subhalo, we take the difference between its most massive match ($M_{\rm{hydro\; subhalo,1}}$) and the sum of any remaining matches ($\sum_{i=2}^{N}M_{\rm{hydro\; subhalo},\mathit{i}}$), and then normalise by $M_{\rm{hydro\; subhalo,1}}$. The PDFs for the fiducial resolution encompass all 40 regions, whereas the lower resolutions are from Region 00 only.}
	\label{fig:mass_ratio_matches}
\end{figure}

Figure \ref{fig:mass_ratio_matches} shows the probability distribution for the differences in mass between the most most massive hydrodynamical subhalo matched to each DMO subhalo and the summed mass of all other matches, normalised by the mass of the most massive match. In all cases except $N/512$ resolution, the mass of the most massive match dominates.

\label{sec:appendix_mass_ratios}
\section{Results on variations to the data pre-processing}
The mean $R^2$ score taken from 1,000 bootstrapped samples of the test set worsens by just 0.002 for both stellar mass and SFR, and by 0.001 for metallicity, when we do not normalise the data by their expected value as a function of subhalo mass during pre-processing, indicating that this pre-processing step makes a negligible difference to machine accuracy. The mean $R^2$ score for galaxy size, meanwhile, remains unchanged. It is to be expected that this normalisation would make little to no difference to galaxy size predictions, due to its much weaker correlation with subhalo mass. However, we can see from the results for stellar mass, SFR, and metallicity, that the models' decisions are not unduly dominated by the most important feature.

We found that the use of weights (dependent on subhalo mass and redshift) in the computation of the loss function during training made no difference to the overall performance metrics of the models. To investigate whether predictions for the rare, high-mass, high-redshift subhaloes were improved by the use of these weights, we binned our data according to subhalo mass and redshift and computed the $R^2$ score in each bin. There is some evidence of improvements made in SFR predictions (on the order of 0.010 in $R^2$) in the higher mass bins, but this "improvement" lies within the bounds of the uncertainties. Crucially, model performance where the bulk of the data lie (lower subhalo masses at $z=5$) is not worsened by our use of training weights. We interpret these results as arising from two conditions: 1) there are few high-mass subhaloes at $z=10$, thus the large weights of these subhaloes do not unduly affect the machine's decision making on low-mass $z=5$ subhaloes, and 2) there tends to be less random variance in the highest-mass galaxies (c.f. Figure \ref{fig:rmsd_rand_seed}), making their properties easier to predict.
\label{sec:appendix_variations}
\section{Correlations in the intrinsic scatter of galaxy properties}
\begin{figure}
    \includegraphics[width=0.5\textwidth]{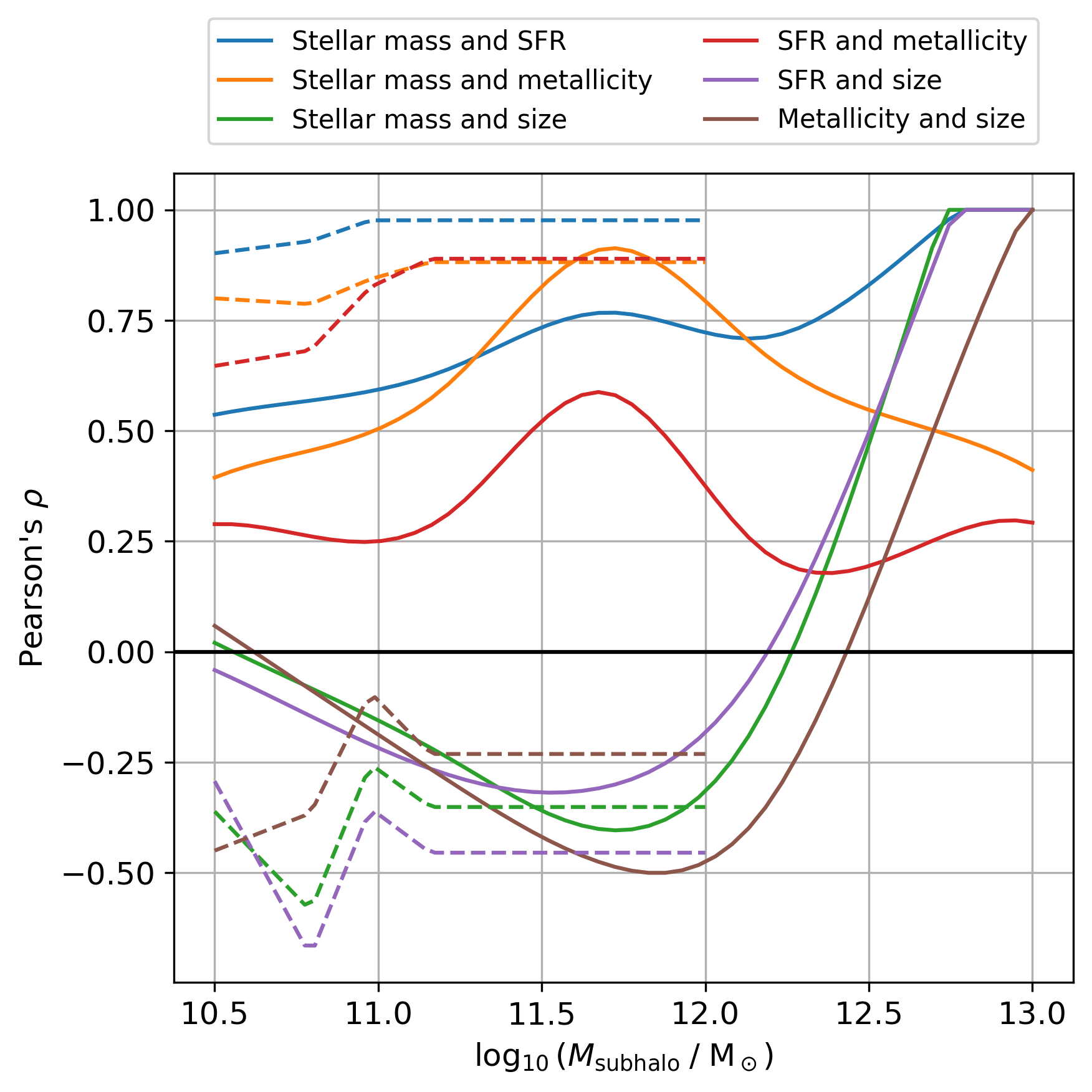}
	\caption{The Pearson's correlation coefficient, $\rho$, for the scatter in each pair of galaxy properties, where we measure the scatter to be the logspace difference between values from two hydrodynamical simulations with different random number seeds matched to the same DMO-simulation subhaloes. The solid lines show the correlations at $z=5$; the dashed lines at $z=10$.}
	\label{fig:scatter_correlations}
\end{figure}

For each pair of galaxy properties, Figure \ref{fig:scatter_correlations} shows the Pearson correlation coefficient, $\rho$, for the pair's logspace differences in Sim A and Sim B. Pearson's $\rho$ must be bound to values in the range [-1,1]. Therefore, we capped instances where the cubic spline would have extended beyond 1.

It can be seen that whenever the stellar mass of a galaxy matched to a given DMO subhalo is larger in one hydrodynamical simulation than it is in the other, the SFR and metallicity tend to be larger in that simulation too, whereas the reverse is true for the stellar HMR at all but the highest subhalo masses.
\label{sec:appendix_scatter_correlations}

%%%%%%%%%%%%%%%%%%%%%%%%%%%%%%%%%%%%%%%%%%%%%%%%%%

% Don't change these lines
\bsp	% typesetting comment
\label{lastpage}
\end{document}